\documentclass[twocolumn]{aastex631}

\usepackage{multirow}
\usepackage{xcolor}
\usepackage{enumitem}

\shorttitle{size-mass relations for DESI galaxies at $z < 0.5$}
\shortauthors{Chen et al.}

\graphicspath{{./}{figures/}}

\begin{document}
	
    \title{Galaxy Clusters from the DESI Legacy Imaging Surveys. \uppercase\expandafter{\romannumeral2}. Environment effects on size-mass relation}
	
    \author{Zhaobin Chen} 
    \affiliation{School of Physics and Technology, Nanjing Normal University, No. 1,
	Wenyuan Road, Nanjing, 210023, P. R. China;
    yuanqirong@njnu.edu.cn}
	
    \author{Yizhou Gu}
    \affiliation{Department of Astronomy, School of Physics and Astronomy, and Shanghai Key Laboratory for Particle Physics and Cosmology, Shanghai Jiao Tong University, Shanghai 200240, P. R. China}
	
    \author{Hu Zou}
    \affiliation{Key Laboratory of Optical Astronomy, National Astronomical Observatories, Chinese Academy of Sciences, Beijing 100012, P. R. China}

    \author{Qirong Yuan}
    \affiliation{School of Physics and Technology, Nanjing Normal University, No. 1,
	Wenyuan Road, Nanjing, 210023, P. R. China; yuanqirong@njnu.edu.cn}
 
    \correspondingauthor{Qirong Yuan}
    \email{yuanqirong@njnu.edu.cn}
	
    \begin{abstract}
		
    To investigate the environment effects on size growth of galaxies, we study the size-mass relation across a broad range of environment with a vast sample of approximately 32 million galaxies at $z < 0.5$ from the DESI Legacy Imaging Surveys. This sample is divided into 3 subsamples representing galaxies within three different environments: brightest cluster galaxies (BCGs), other cluster galaxies, and field galaxies.
    The BCGs in our large sample are dominated by quiescent galaxies (QGs), while only a minority ($\thicksim$13\%) of BCGs are star-forming galaxies (SFGs). 
    To demonstrate the influence of environment on size growth, we attempt to observe the difference in size-mass relation for these three subsamples. 
    In general, the slope differences between QGs and SFGs within various environments are significant, and tend to be greater at higher redshifts. 
    For the mass-complete subsamples at $z<0.5$, BCGs are found to have the highest slope of size-mass relation, and no difference in size-mass relation is found between cluster members and field galaxies. 
    To assess whether the observed slope differences stem from the variations in environment or mass distribution, we construct the mass-matched subsamples for QGs and SFGs. As a result, both QGs and SFGs show negligible difference in slope of size-mass relation among the galaxies within three distinct environments, indicating that stellar mass is the most fundamental factor driving the size evolution at $z < 0.5$, though the mass growth mode for QGs and SFGs may have been affected by galaxy environment. 
		
    \end{abstract}   
    \keywords{\href{https://vocabs.ands.org.au/repository/api/lda/aas/the-unified-astronomy-thesaurus/current/resource.html?uri=http://astrothesaurus.org/uat/594} {Galaxy evolution (594)}; \href{https://vocabs.ands.org.au/repository/api/lda/aas/the-unified-astronomy-thesaurus/current/resource.html?uri=http://astrothesaurus.org/uat/2029}{Galaxy environments (2029)}; \href{https://vocabs.ands.org.au/repository/api/lda/aas/the-unified-astronomy-thesaurus/current/resource.html?uri=http://astrothesaurus.org/uat/181}{Brightest cluster galaxies (181)}
    }
		
    \section{Introduction} \label{sec:introduction}

    In the standard hierarchical model, weak density perturbations in the early Universe lead to the gradual collapse of dissipationless dark matter into dense dark matter halos (\citealt{1998MNRAS.295..319M}). Within these halos, primitive gas gradually cools to form gaseous nebulae, which, under the influence of gravitational potential wells, collapse into dense, luminous nuclei, or galaxies (e.g., \citealt{1978MNRAS.183..341W,1991ApJ...379...52W}). As galaxies evolve, collisions and mergers occur, resulting in the formation of larger galaxies. Throughout this process, the angular momentum provided by dark matter halos dominates the size of galaxies (e.g., \citealt{1998MNRAS.295..319M,2013ApJ...764L..31K,2017ApJ...838....6H}). Thus, tracking the coevolution of galaxy size and mass with cosmic time provides insight into the assembly history of galaxies.  

    Stellar mass, as an important factor in the growth of galaxy size, has been extensively studied (e.g., \citealt{2003MNRAS.343..978S,2010ApJ...713..738W,2015ApJS..219...15S}). 
    Stellar mass assembly is a crucial process in galaxy evolution, which is influenced by intricate factors, such as star formation and mergers (e.g., \citealt{2010ApJ...709..644I,2015ApJ...814...96W,2010ApJ...709.1018V}). Usually, massive galaxies form earlier than lower-mass galaxies (e.g., \citealt{2018ApJ...855...10G,2022ApJ...934..177K}), although some studies on individual galaxies suggest somewhat different histories of stellar mass assembly (e.g., \citealt{2022MNRAS.516.3318W}). 
    According to \cite{2012ApJ...747...34B}, mergers are considered as the predominant factor driving the observed size evolution in massive galaxies below $z\thicksim 3$, resulting in a threefold increase in mass from $z = 3$ to the present day solely through merging. Numerous investigations suggest that minor mergers may have a noteworthy impact on the growth of quiescent galaxies (e.g., \citealt{2007ApJ...658..710N, 2012ApJ...746..162N, 2012ApJ...750...32G, 2015ApJ...813...23V, 2019ApJ...887..113W}). In addition, \cite{2013MNRAS.428.1088M} reports compelling evidence supporting the notion that minor mergers play a crucial role in the evolution of massive galaxies at $1.3 < z < 1.5$. 
    \cite{2010ApJ...709.1018V} found that the mass in the central regions of massive galaxies remains roughly constant with redshift, and their growth since $z \sim 2$ is only due to the gradual increase in their outer envelopes. 
    
    \cite{2009ApJ...691.1879W} suggest that the bimodal distribution of galaxies in color space can be traced back to $z \sim 2$. 
    In general, star-forming galaxies (SFGs) with a bluer color, whose mass growth is dominated by star formation, become quiescent galaxies (QGs) with a redder color after quenching (e.g., \citealt{2014MNRAS.440..889S, 2017MNRAS.471.2687B, 2019ApJ...884..172G}).  It is widely believed that the star formation activity in galaxies is closely related to their morphologies  (\citealt{2018MNRAS.481.3456C,2019A&A...624A..80N}).
    Morphologically speaking, SFGs with disc-dominated morphologies, which are inclined to late-type galaxies, are considered to be growing inside-out through continuous star formation (e.g., \citealt{2011MNRAS.418.2493P,2013ApJ...773...43B}). However, QGs are inclined to early-type galaxies with elliptical or bulge-dominated morphologies. According to the work of \cite{2023A&A...669A..11P}, only 8\% of late-type galaxies are quiescent.  
   
    The correlation between galaxy morphology and environmental density has long been established (e.g., \citealt{1974ApJ...194....1O,1980ApJ...236..351D}). 
    The divergence in the environmental distribution of blue and red galaxies appears since $z < 1.5$ (\citealt{2018ApJ...855...10G}).
    Field and cluster galaxies represent two distinct populations residing in markedly different environments, and extensive research has been conducted to explore the processes of star formation and mass assembly within each of these populations (e.g., \citealt{2006ApJ...651..120B,2023ApJ...955...32K}). From a theoretical perspective, cluster galaxies in denser environment are suffering more effects of the dense and hot intracluster medium than field galaxies (e.g, \citealt{1972ApJ...176....1G,2003ApJ...596L..13B}). Indeed, \cite{2017ApJ...834...73Y} find that local early-type galaxies in the densest environment with stellar mass greater than $ 10^{11.2} M_\odot$ tend to have larger sizes than those in low-density environment.
    While some researchers have suggested that cluster galaxies may be slightly smaller when compared to their counterparts in the field (e.g., \citealt{2013ApJ...762...77P,2014MNRAS.444..682C}). 
    Despite the different conclusions reached by these studies, they all share the common idea that galaxy morphology is closely related to the environment. Nevertheless, \cite{2017A&A...597A.122S} demonstrate that at approximately $z$ = 1.3, the structures and intrinsic properties (namely, the equivalent median effective radius and radius normalized mass) of cluster and field elliptical galaxies are comparable. Similarly, \cite{2009MNRAS.394.1213W} find that no differences in radius are observed between early-type central and satellite galaxies. The analysis of \cite{2013ApJ...779...29H} also indicates no significant environmental dependence for the sizes of central and satellite early-type galaxies at fixed stellar mass at $z$ $\thicksim$ 0. 
 
    Among the member galaxies of cluster, the brightest cluster galaxies (BCGs) are expected to have a more complex evolution process, due to the association with dense environment at the center of cluster. Generally, BCGs are found close to the center of galaxy clusters with the densest environment, which has been determined by X-ray or gravitational lens observations (e.g., \citealt{1984ApJ...276...38J,2005MNRAS.359..417S}). Compared with member and field galaxies, BCGs may have a more complex pattern of mass growth (e.g., \citealt{2020ApJS..247...43K}). Simulations and semi-analytic models show that BCGs assemble most of their stellar mass through multiple mergers (e.g., \citealt{1998ApJ...502..141D,2007MNRAS.375....2D}). Observational studies also suggest that merger plays an important role in the evolution of BCGs (e.g., \citealt{2013MNRAS.433..825L,2016MNRAS.462.4141L}). 
    Further investigations on the stellar mass evolution of BCGs show that their mass increases as redshift decreases (e.g., \citealt{2012MNRAS.427..550L}), while others maintain that the stellar mass has remained constant since at least $z \thicksim 1 $ (e.g., \citealt{2008MNRAS.387.1253W,2011ApJ...726...69A}).
    The work of \cite{2015ApJ...814...96W} demonstrates that the evolutionary pattern of stellar mass varies with redshift. Specifically, when redshifts exceed 1, the dominant force driving the growth of stellar mass may be the process of star formation, whereas at lower redshifts ($z<1$), dry mergers are likely to play a more significant role in this phenomenon. Some studies suggest that the growth rate of BCG mass may exhibit a slowdown, with more rapid growth observed at $z > 0.5$ and slower growth at $z < 0.5$ (e.g., \citealt{2013ApJ...771...61L,2014MNRAS.440..762O}). 
    In addition, diverse perspectives in research provide unique insights, such as the study by \citet{2011ApJ...726...69A} on the evolution of BCGs over the past approximately 6 Gyr. They believe that the only mechanism that is able to explain the evolution in BCG size and the non-evolution in the BCG S$\rm \acute{e}$rsic shape parameter during the last 6 Gyrs are the feedback processes (such as galactic winds and ejection).  

    Taking into account the environment as a variable in our study, we attempt to investigate whether it plays a significant role in the variation of the size-mass relation ($R_{\rm e} \propto M_*^{\alpha}$). In this paper, we select three subsamples for our analysis, including field galaxies, member galaxies, and BCGs. Based on the ninth publicly released data of the Dark Energy Spectroscopic Instrument (DESI) Legacy Imaging Surveys, we can obtian more than 32 million galaxies with a sky area of $\thicksim$ 20,000 deg$\rm ^2$ in the redshift range of $0< z < 0.5$. The influence of environment on size-mass relation is analyzed. This paper is the second in our series of works (the first study has already been completed by \citealt{2021ApJS..253...56Z}) and is organized as follows. 
    In Section \ref{sec:data and sample}, we describe the catalogue of data from the DESI Legacy Imaging Surveys and the detailed process to construct samples. In Section~\ref{sec:size-mass relation}, we analyze the detailed influence of stellar mass and environment on the size-mass relation. 
    Section~\ref{sec:discussion} presents our discussions about the dependence of the size-mass relation upon environment and redshift.  
    Summary and conclusion are presented in Section \ref{sec:conclusion}. Throughout our paper, we adopt the following cosmological parameters: $\rm \Omega_m$ = 0.30, $\rm \Omega_{\Lambda}$ = 0.70, and $\rm H_0$ = 70 km $\rm s^{-1}$ $\rm Mpc^{-1}$

    \vspace{10pt}
    \section{Data and Sample Selection} \label{sec:data and sample}
    
    \subsection{Data Description} \label{sec:2.1}
 
    The ninth public data release (DR9) of the Legacy Imaging Surveys of DESI covers about 20,000 square degrees in both the South and North Galactic Caps, which provides imaging in $g/r/z$ bands with 5$\sigma$ depth of 24.7/23.9/23.0 (e.g., \citealt{2019AJ....157..168D,2022RAA....22f5001Z}). It consists of three independent optical surveys using three different telescopes: the Beijing-Arizona Sky Survey (BASS; \citealt{2017PASP..129f4101Z}), Mayall $z$-band Legacy Survey (MzLS), and DECam Legacy Survey (DECaLS). In addition, the imaging surveys provide near-infrared photometric data in the 3.4 and 4.6 $\mu m$ WISE bands (i.e., Wide-field Infrared Survey Explorer $W1$ and $W2$, see \citealt{2010AJ....140.1868W}). This data release also includes the 6-year imaging of WISE, with 5$\sigma$ depths of 20.7 and 20.0 in $W1$ and $W2$ bands, respectively. 
	
    \cite{2022RAA....22f5001Z} expand their prior research on photometric redshift and cluster detection to the DR9 of DESI Legacy Imaging Surveys, generating a sample of galaxy clusters over 540,000 in the redshift range of $0 < z < 1$. We start from the magnitude-limited sample to identify the galaxy clusters in Section \ref{sec:2.3}, using the following criteria listed in \cite{2022RAA....22f5001Z} on the DR9 of the legacy imaging surveys:
	\begin{enumerate}
		\item $r <$ 23 (the upper limit of $r$-band magnitude for excluding faint sources). 
		\item type != ``PSF" (the ``PSF" refers to point source). 
		\item fracmasked\_[$g,r,z$] $<$ 0.5 (clean photometric cuts using the profile-weighted fraction of pixels masked).
		\item fracflux\_[$g,r,z$] $<$ 0.5 (clean photometric cuts using the profile-weighted fraction of the flux).
		\item fracin\_[$g,r,z$] $>$ 0.3 (clean photometric cuts using the fraction of a source's flux within the blob). 
		\item it is detected in the five bands ($g, r, z, W1, W2$). 
		\item the photo-$z$ error is set to less than 0.1$\times$(1 + $z_{\rm phot}$). 
		\item the range of the $r$-band absolute magnitude is
		$-25 < M_r < -16$. 
		\item the stellar mass range is $10^6  < M_*/M_\odot < 10^{13}$. 
		\item the uncertainty of logarithmic stellar mass $|\Delta \log(M_*/M_\odot)| < 0.4$ dex. 
	\end{enumerate}
	The photometric redshifts, stellar masses, colors and star formation rates (SFRs) are taken from the catalogue of \cite{2019ApJS..242....8Z}, which is updated to the DESI DR9 release (\citealt{2022RAA....22f5001Z}).
	
    Using a hybrid technique to estimate photometric redshift ($z_{\rm photo}$), a local linear regression algorithm is applied to a dedicated spectroscopic training set, and then a spectrum template is expertly fitted to determine $k$-corrections and absolute magnitudes \citep{2016MNRAS.460.1371B, 2018ApJ...862...12G}. The stellar mass ($M_{\rm *}$) and SFR are derived by fitting the photometric spectral energy distribution (SED) of galaxies using the LePhare software \footnote{\url{https://www.cfht.hawaii.edu/˜arnouts/LEPHARE/lephare.html}} (\citealt{2002MNRAS.329..355A,2006A&A...457..841I}). The default \cite{2003MNRAS.344.1000B} (BC03) spectral models are employed, in conjunction with the initial mass function (IMF) from \cite{2003PASP..115..763C}. Three metallicities (0.004, 0.008, and 0.02) and an exponentially declining star formation history (SFH) from 0.1 to 30 Gyr are taken. For more detailed photometric redshift and stellar mass information please refer to \cite{2019ApJS..242....8Z}.	
	
    Throughout the paper, the size of galaxy is determined by the half-light radius ($R_{\rm e}$) of galaxy model with \textit{the Tractor}. \textit{The Tractor} (\href{http://www.ascl.net/1604.008}{Lang et al. 2016}) performs profile-fitting photometry. By assuming the same model across all the bands, model fits are determined using only the optical $g/r/z$ data with the specific point spread function (PSF). Four types of the galaxy profiles are used in the fitting procedure of DR9: round exponential galaxies with a variable radius (``REX''), de Vaucouleurs (``DEV'') profiles (elliptical galaxies), exponential (``EXP'') profiles (spiral galaxies), and S\'ersic (``SER'') profiles. The best-fit model is determined by convolving each model with the specific PSF for each band and minimizing the residuals.

    \subsection{Galaxy Clusters} \label{sec:2.2}
 
    Previously, \cite{2020PASP..132b4101G} developed a rapid clustering algorithm called Clustering by Fast Search and Find of Density Peaks (CFSFDP) to identify galaxy clusters and their members from the photometric redshift catalog.
    The key aspect of this algorithm lies in providing a simple and efficient approach for cluster identification, which can quickly identify galaxy cluster centers exhibiting higher density compared to their neighboring regions. This approach successfully eliminates the distorting influence of foreground and background galaxies. Using this algorithm on their sample, \cite{2021ApJS..253...56Z} identify 540,432 galaxy clusters at z $\lesssim$ 1 from the DESI legacy imaging surveys. As the galaxy closest to the position of density peak is not always the BCG, they identified the $r$-band brightest galaxy within 0.5 Mpc around the density peak as the BCG.
	
\begin{figure}[ht!]  
            \includegraphics[width=1.\columnwidth]{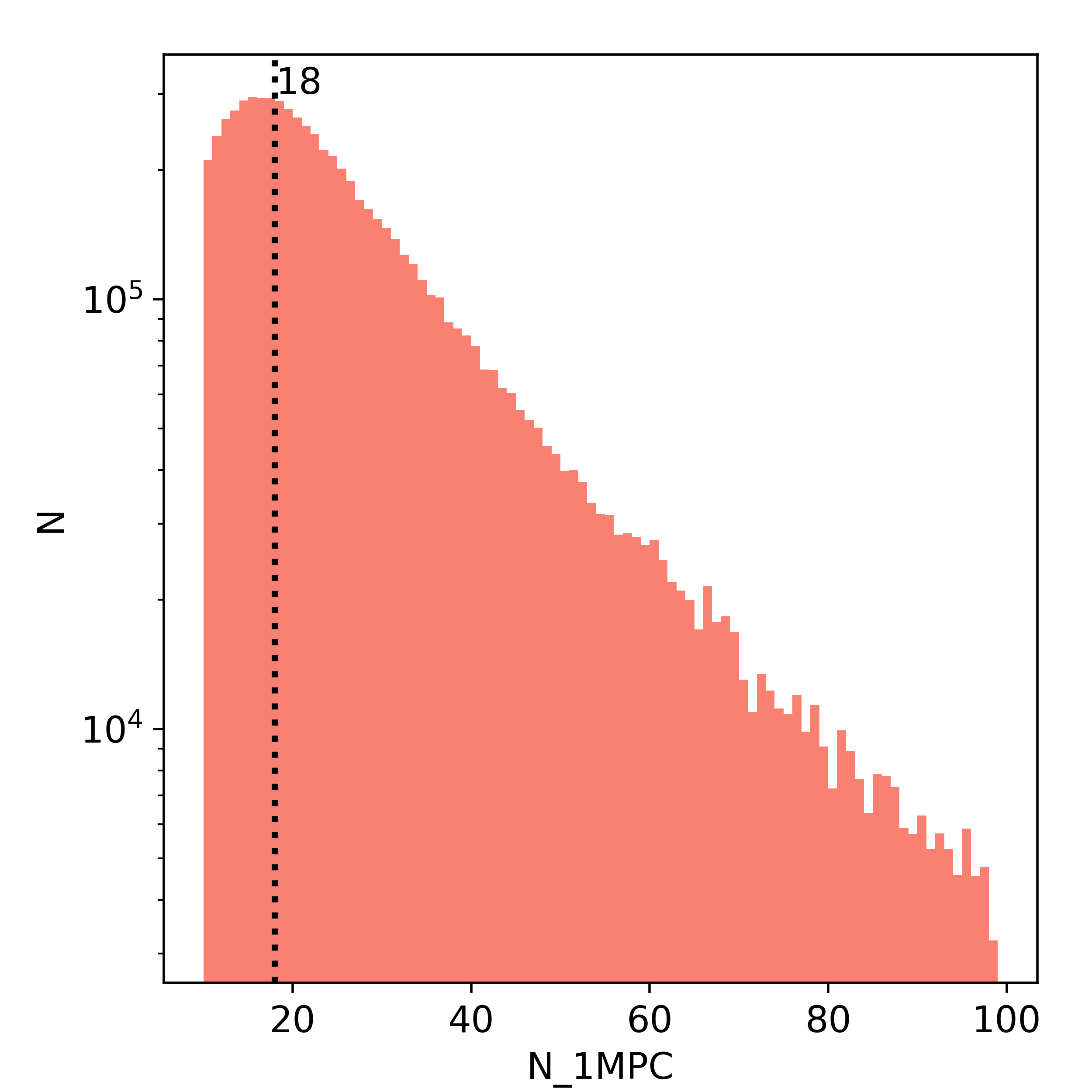}
            \caption{Illustrated in the figure is the distribution of galaxy cluster richness, where $N_{1\hspace{0.3em}\rm Mpc}$ signifies the richness of the cluster. The vertical black dashed line denotes the peak of the distribution at $\sim 18$. 
            \label{fig:N-1MPC}}
        \end{figure}

    For each galaxy cluster, the number of member galaxies within a distance of 1 Mpc from its center ($N_{1\hspace{0.3em}\rm Mpc}$) is used as an estimate of the cluster richness, after correcting for the number of galaxies expected from the background density. Figure \ref{fig:N-1MPC} displays the histogram distribution of the richness $N_{1\hspace{0.3em}\rm Mpc}$ for galaxy clusters. The distribution peaks at around 18, which suggests that galaxy clusters with $N_{1\hspace{0.3em}\rm Mpc} > 18$ can be considered as rich clusters with high completeness. 

    The catalogs of photometric redshifts and clusters used in this paper are available through the ScienceDB Web site\footnote{\url{https://www.doi.org/10.11922/sciencedb.o00069.00003}}.
  
    \subsection{Sample Selection}  \label{sec:2.3}
 
	We select three distinct subsamples from the galaxy cluster catalog, representing three different environments. These subsamples include BCGs located in the densest environmental region of the cluster; member galaxies inhabiting regions of environmental density second only to BCGs; and field galaxies isolated within low-density environments. Our approach to constructing these subsamples is elaborated below:
 
        \begin{enumerate}
            \item We start with the sample from \cite{2022RAA....22f5001Z}, selecting galaxies at $z < 0.5$. The upper limit of redshift 0.5 is ultimately determined by the fact that the corresponding physical size for a typical point spread function in the $r$-band images ($\sim 0.''8$) is about 5 kpc at $z=0.5$. If spectroscopic redshift ($z_{\rm spec}$) is available, we use it preferentially; Otherwise, photometric redshift would be used. 
            \item To form a subsample of field galaxies, we removed all cluster galaxies, including both BCGs and member galaxies.
            \item Taking into account the completeness of galaxy cluster, clusters with a richness greater than 18 (i.e., $N_{1\hspace{0.3em}\rm Mpc}$ > 18) are selected. Once the clusters are selected, we extract the corresponding BCGs and member galaxies to construct the subsamples of BCGs and member galaxies. To reduce the degree of contamination in membership, we select the galaxies within a projected distance of 0.5 Mpc from the BCGs as member galaxies.
        \end{enumerate}

    To further investigate the size-mass relation in various types of galaxies, we categorize the galaxies into two types: QGs and SFGs. Numerous studies have shown that QGs and SFGs represent two distinct stages of galaxy evolution, as evidenced by the bivariate distribution in color space (e.g., \citealt{2004ApJ...600..681B,2014MNRAS.440..889S}). SFGs are a type of galaxies with relatively high SFRs and active star-forming activities. On the other hand, QGs have entered a phase of inactivity with the cessation of star formation, exhibiting lower SFRs. As reported by \cite{2014SerAJ.189....1S} and \cite{2017MNRAS.471.2687B}, there is a remarkable correlation between morphology and specific star formation rate ({\rm sSFR} = SFR/$M_{*}$). 
    
\begin{figure*}[ht!]  
            \includegraphics[width=\textwidth]{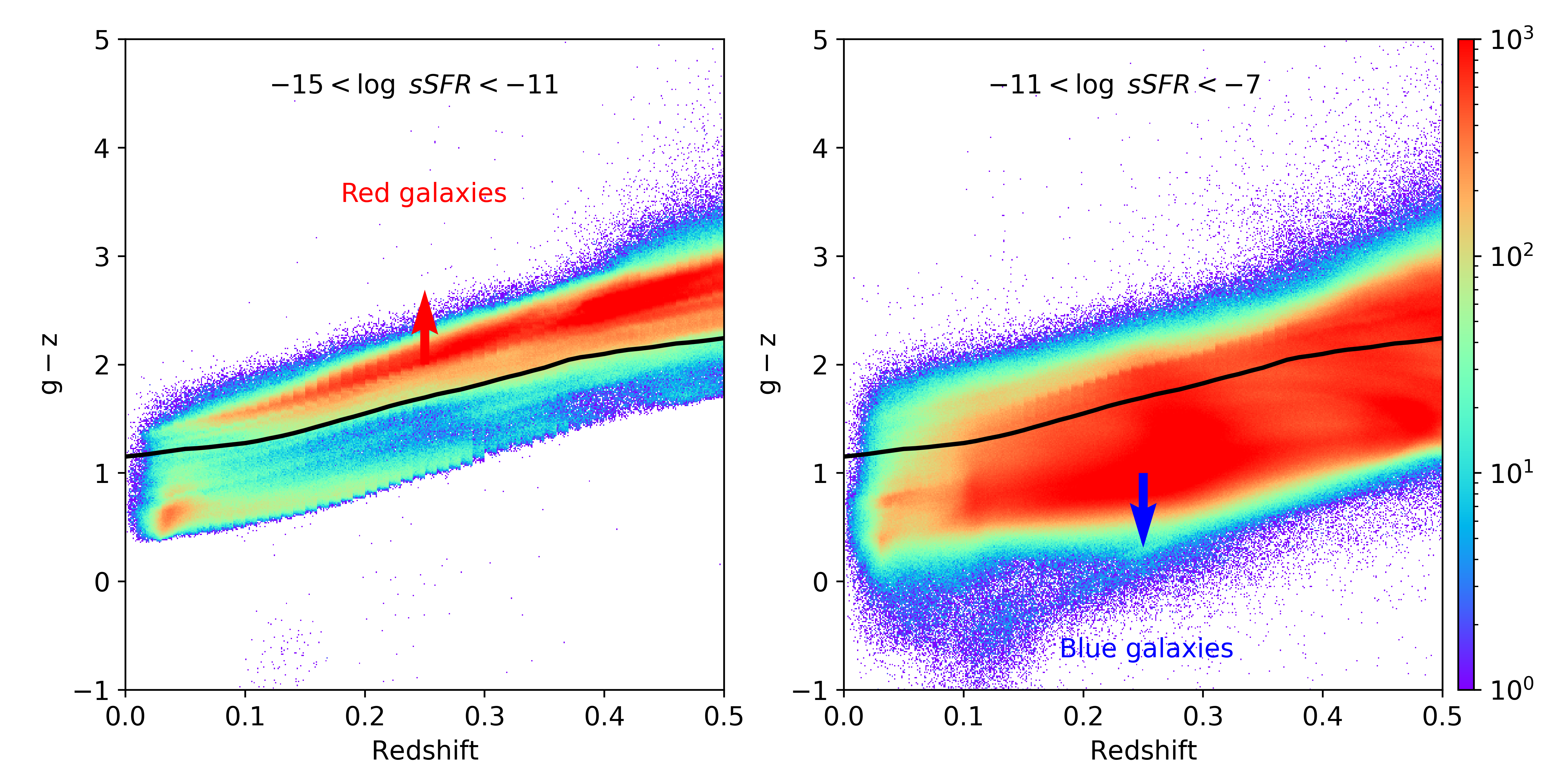}
                \caption{Rest-frame color-redshift diagram for galaxies with redshifts in the range of $0 < z < 0.5$. The solid black line represents the BC03 template with $\tau = 2$ Gyr. Left panel: Distribution of the sample of quiescent galaxies, which are selected using the {\rm sSFR} criterion (-15 < $\log {\rm sSFR}$ < -11). Galaxies above the solid black line are classified as QGs. Right panel: Distribution of the sample of star-forming galaxies, which are selected using the {\rm sSFR} criterion (-11 < $\log {\rm sSFR}$ < -7). Galaxies below the solid black line are classified as SFGs. The color bar represents the number of galaxies.
            \label{fig:r-b-galaxies}}
        \end{figure*}

    It is essential to provide a combined approach that can distinguish between QGs and SFGs more effectively. As early as the work by \cite{2010MNRAS.406.2249W}, researchers select ``passive galaxies" using the criterion $\log {\rm sSFR} < -11$. 
    Subsequently, an increasing number of studies adopt the criterion of $\log {\rm sSFR} = -11$ to differentiate between these two galaxy populations (e.g., \citealt{2018ApJ...859...11S,2023MNRAS.524.5314K}). 
    However, considering the uncertainties in determining stellar mass and SFR, the selection alone by {\rm sSFR} criterion may not be sufficient.  
    As shown in Figure \ref{fig:r-b-galaxies},  many SFGs selected by $-11 < \log {\rm sSFR} < -7$ appear to be red.
    Leveraging the advantage of the large sample size provided by the DESI Legacy Imaging Surveys, we can employ more stringent criteria to select QGs and SFGs more reliably. 
    The study conducted by \cite{2013A&A...556A..55I} reveals that, at low-redshift ($z < 1$), the independent selection of QGs using the criteria of {\rm sSFR} ($\log {\rm sSFR} < -11$) or color demonstrates a high level of concordance. 
    In practice, we use both {\rm sSFR} and color as selection criteria to categorize galaxy populations. 
    As a result, after applying the color criterion described below, about 22\% red galaxies are found to have $-11 < \log {\rm sSFR} < -7$, and about 7.6\% blue galaxies have $-15 < \log {\rm sSFR} < -11$. 
    
    Specifically, we have employed five criteria, as listed below: 
    \begin{enumerate}
        \item The $r$-band magnitude less than 22 to ensure the quantity of photometric redshift (see appendix \ref{sec:appendix}). 
        \item Considering the reliability of the half-light radius measurement, the galaxies with $R_{\rm e} / \sigma_{R_{\rm e}} > 5$ are selected. 
        \item A more rigorous criterion than parent catalog is applied to the photometric redshifts, $\sigma_{z_{\rm photo}} < 0.05 \times(1+z_{\rm photo})$, to exclude the galaxies with poor SED fittings or inaccurate redshift estimates.  
        \item The specific star formation rate ({\rm sSFR}) is utilized to distinguish between quiescent and star-forming galaxies, with the former having  $-15<{\log {\rm sSFR}}<-11$ and the latter with of $-11<{\log {\rm sSFR}}<-7$.	
        \item The color criterion  is further taken to distinguish the SFGs and QGs. Based on the BC03 templates, the color cut of $(g-z)$ as a function of redshift can be established by tracing the galaxy template with an exponentially declining SFH ($t = 5\hspace{0.3em} \rm Gyr; Z_\odot; \tau < 2\hspace{0.3em} \rm Gyr $). We select red and blue galaxies from the quiescent and star-forming populations, respectively, and form a final sample consisting of both types in three distinct environments. Figure \ref{fig:r-b-galaxies} visually illustrates the process of sample selection. 	
    \end{enumerate}

	\begin{deluxetable}{c|ccc|c}[ht!]   
		\tablenum{1}
		\tablecaption{Sample size statistics for different environments
		\label{tab:1}}
		\tabletypesize{\scriptsize}
		\tablewidth{0pt}
		\tablehead{ \multicolumn{1}{c|}{Type} & \colhead{Field} & \colhead{Member} & \multicolumn{1}{c|}{BCGs} & \colhead{Sum} }
		\startdata
		QGs & 6,844,819 & 461,636 & 29,827 & 7,336,282
		\\
		(percentage) & (21.36\%) & (1.44\%) & (0.10\% ) & (22.90\%)
            \\
		SFGs & 24,335,569 & 363,014 & 4,495 & 24,703,078 
		\\
		(percentage) & (75.96\%) & (1.13\%) & (0.01\% ) & (77,10\%)
		\\
		\hline
		Sum & 31,180,388 & 824,650 & 34,322 & 32,039,360 
		\\
		(percentage) &(97.32\%) & (2.57\%) &(0.11\% ) & (100\%)
		\\
		\hline
		Total &  \multicolumn{3}{c|}{32,039,360} & /
		\\
		\enddata
		\tablecomments{The categories``Field, Member, BCGs, QGs, SFGs" correspond to field galaxies, member galaxies, the brightest cluster galaxies, quiescent galaxies, and star-forming galaxies, respectively. The percentages indicate the fraction of different types of subsamples to the total sample.}	
	\end{deluxetable}
    
        \begin{deluxetable}{cccccc}[ht!]  
        \tablenum{2}
        \tablecaption{The statistics of BCGs in different redshift bins
        \label{tab:2}}
        \tabletypesize{\scriptsize}
        \tablewidth{0pt}
        \tablehead{ \colhead{Redshift} & \colhead{$0-0.1$} & \colhead{$0.1-0.2$} & \colhead{$0.2-0.3$} & \colhead{$0.3-0.4$} & \colhead{$0.4-0.5$} }
        \startdata
        $\rm N_{QG}$ & 601 & 4,626 & 8,510 & 7,372 & 8,718 
        \\
        percentage & 2.01\% & 15.51\% & 28.53\% & 24.72\% & 29.23\% 
        \\
        \hline
        $\rm N_{SFG}$ & 100 & 352 & 946 & 1,002 & 2,095
        \\
        percentage & 2.22\% & 7.83\% & 21.05\% & 22.29\% & 46.61\% 
        \\ \hline
        $\rm N_{ALL}$ & 701 & 4,978 & 9,456 & 8,374 & 10,813 
        \\
        percentage & 2.05\% & 14.50\% & 27.55\% & 24.40\% & 31.50\% 
        \\ \hline
        $\rm \Sigma (N_{QG})$ & \multicolumn{2}{|c|}{29,827 (86.90\%)} & $\rm \Sigma (N_{SFG})$ & \multicolumn{2}{|c|}{4,495 (13.10\%)}
        \\
        \enddata
        \tablecomments{``$\rm N_{QG}$" represents the number of quiescent galaxies in the redshift bin, while ``$\rm N_{SFG}$" represents the number of star-forming galaxies and ``$\rm N_{ALL}$" represents the total number of galaxies. ``$\rm \Sigma (N_{QG})$" and ``$\rm \Sigma (N_{SFG})$" represents the total number of quiescent and star-forming galaxies without bin distinction, respectively.}		
        \end{deluxetable}

	\begin{deluxetable}{ccccc}     
		\tablenum{3}
		\tablecaption{The complete mass at different magnitude limits in different redshift bins
		\label{tab:3}}
		\tabletypesize{\scriptsize}
		\tablewidth{0pt}
		\tablehead{ \colhead{Redshift} & \colhead{Type} & $r_{\rm lim}$ & \colhead{$N$} & \colhead{log$(M_{\rm *,comp}/M_\odot) $} }
		\startdata
		$0-0.1$   &  QGs & 18.4 & 83,136 & 10.12 
		\\
		$0.1-0.2$ &  QGs & 19.7 & 508,080 & 10.32
		\\
		$0.2-0.3$ &  QGs & 20.5 & 1,194,350 & 10.41
		\\
		$0.3-0.4$ &  QGs & 21.4 & 1,652,403 & 10.29
		\\
		$0.4-0.5$ &  QGs & 22.0 & 2,340,092 & 10.28
		\\
		\hline
		$0-0.1$   &  SFGs & 22.0 & 809,587 & 8.10 
		\\
		$0.1-0.2$ &  SFGs & 22.0 & 3,705,206 & 8.79
		\\
		$0.2-0.3$ &  SFGs & 22.0 & 7,414,001 & 9.21
		\\
		$0.3-0.4$ &  SFGs & 22.0 & 4,391,249 & 9.54
		\\
		$0.4-0.5$ &  SFGs & 22.0 & 3,564,723 & 9.85
		\enddata
		\tablecomments{``$r_{\rm lim}$" represents $r$-band magnitude;  ``$N$" represents the number of galaxies that are greater than the mass-complete  limit in correspondind redshift interval ($\vartriangle z$ = 0.1); ``log$(M_{\rm *,comp}/M_\odot)$" represents the mass-complete  limit adopted for the corresponding redshift bin. }	
	\end{deluxetable}

        \begin{figure*}[p]  
            \begin{minipage}{0.99\textwidth}
                \centering
        	\includegraphics[width=\textwidth]{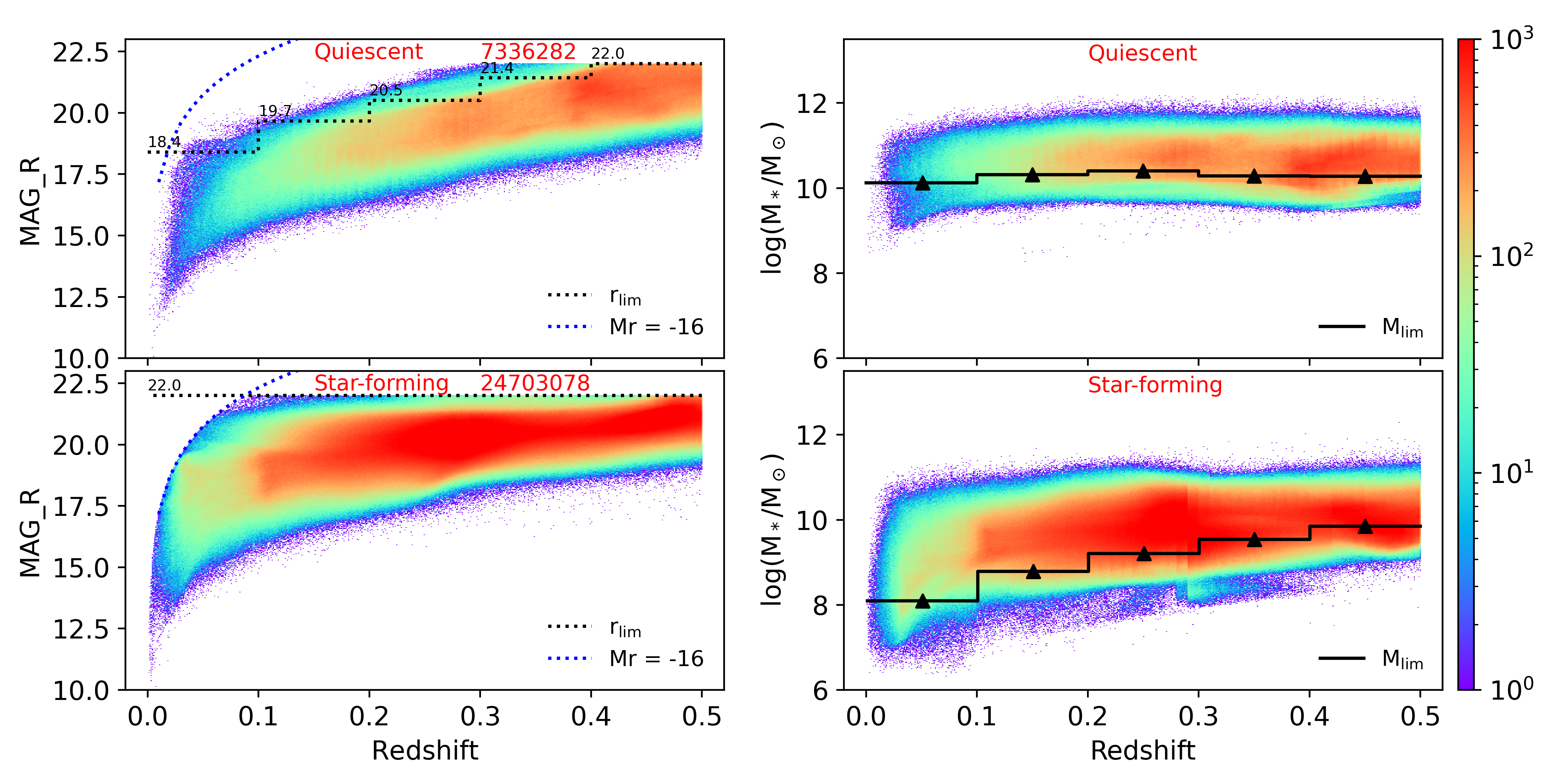}
                \caption{Left: Upper and lower panels display the distribution of $r$-band magnitudes as a function of redshift. Different colors represent different distribution densities of galaxies. Red numbers indicate the total number of galaxies in the subsample. The black dotted lines with black numbers represent the apparent magnitude limit values ($r_{\rm lim}$) in the corresponding redshift bins, while the blue dotted lines display the $r$-band absolute magnitude equal to -16 without considering the $k$-correction. The top panel displays the distribution of QGs with $r_{\rm lim}$ sequentially used as (18.4, 19.7, 20.5, 21.4, 22.0), and the bottom panel illustrates the distribution of SFGs, for which we adopt a fixed $r_{\rm lim}$ value of 22.0 in all redshift bins. The rationale for using different methods to determine the magnitude limits is detailed in section \ref{sec:2.3}. 
                Right: Upper and lower panels present the distribution of stellar masses as a function of redshift. The black solid lines in the panel denote the corresponding completeness of stellar mass ($M_{\rm comp}$) in different bins, while the black triangle indicates the median redshift in each bin. 
        	\label{fig:m-comp}}
            \end{minipage}
            
            \vspace{2\baselineskip}
            
            \begin{minipage}{0.99\textwidth}
                \centering
        	\includegraphics[width=\textwidth]{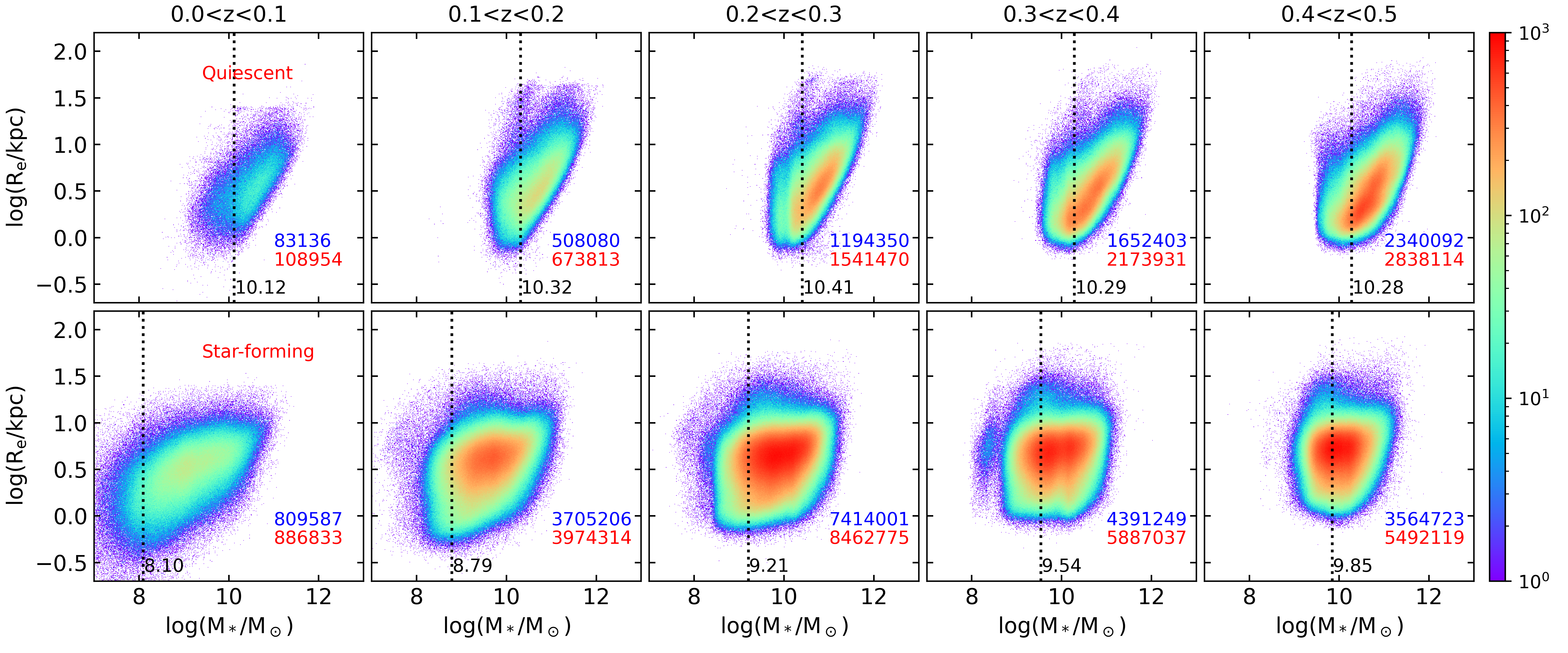}
                \caption{Size-mass distribution map ($R_{\rm e} - M_{\rm *}$) of the total sample. The distribution of $R_{\rm e}$ and $M_{\rm *}$ is shown in logarithmic space. The top five panels show the distribution of QGs in five different redshift bins, while the bottom five panels show the distribution of SFGs. The color gradients in the figure indicate the density of galaxies, and the blue/red number in the lower right corner indicates the number of galaxies with mass-complete/total galaxies in that particular redshift bin. The black dotted lines represent the 90\% mass-complete limits.
        	\label{fig:r-m-comp}}
            \end{minipage}
	\end{figure*}

    Our final total sample comprises 32,039,360 galaxies with $r < 22$ at $z < 0.5$. These galaxies are classified into two types: QGs and SFGs, and have been categorized into three different environments: field galaxies, member galaxies, and BCGs. 
    Statistics on the number of subsamples under different conditions are shown in  Table \ref{tab:1}, in which field galaxies occupy the majority. Additionally, we provide a count distribution of BCGs, showing that star-forming galaxies constitute a very small fraction (about 13.10\%), as demonstrated in Table \ref{tab:2}.

    The 90\% completeness limit of stellar mass (mass-complete: $M_{\rm comp}$) is estimated using a method similar to \cite{2010A&A...523A..13P} and \cite{2021ApJ...921...60G}. The completeness of stellar mass is affected by both the redshift and mass-to-light ratio (M/L ratio) for each galaxy. Given the well-established fact that quiescent galaxies usually have a higher M/L ratio compared to star-forming galaxies, the completeness limits for different galaxy types can vary at different redshifts.
    We calculate the value of $M_{\rm comp}$ for each redshift bin by determining the stellar mass limit ($M_{\rm lim}$) for the 20\% faintest galaxies within a given redshift bin ($\vartriangle z$ = 0.1), by applying the following formula:
    $    \log M_{\rm lim} = \log M_* + 0.4 \times (r - r_{\rm lim}) $, where $r_{\rm lim}$ is set as the magnitude limit of our subsamples.
    For SFGs we use a fixed value of 22.0 as the magnitude limit for each redshift bin.
    For QGs, based on the distribution of magnitudes at $z < 0.4$, we pick up the $r$-band magnitude than which  95\% galaxies are brighter at each redshift bin as the limiting magnitude ($r_{\rm lim}$). As a result, the $r_{\rm lim}$ for the five redshift bins  are determined to be 18.4, 19.7, 20.5, 21.4, and 22.0, respectively. Subsequently, the final $M_{\rm comp}$ is defined as the upper envelope of $M_{\rm lim}$ distribution below which lie 90\% of the $M_{\rm lim}$ values at a given redshift bin.   
		
    Figure~\ref{fig:m-comp} depicts the distribution of $r$-band magnitude and stellar mass as a function of redshift. The imposed limit on the $r$-band absolute magnitude ($M_r < -16$) leads to a dearth of galaxies at the faint end at lower redshifts, as evident from the left panels. However, quiescent galaxies show a different pattern of deficiency compared to star-forming galaxies, with a more pronounced lack of the former at the faint end. After reviewing our sample and selection criteria, we recognize that some fainter galaxies are excluded due to our implementation of more strict criteria on both colors and {\rm sSFR}s. Only the quiescent galaxies with red colors are included. In addition, the lack of faint quiescent galaxies at lower redshift may be also related with  observational constraints and data processing methods. 
	
    Due to this, we adopt the dynamic limits of apparent magnitude to estimate the completeness of stellar mass. The statistics of stellar mass limit and galaxy number in this sample are shown in Table \ref{tab:3}, in which a brighter limit of apparent magnitude is settled at lower redshift. Figure~\ref{fig:m-comp} also illustrates the adopted limit of apparent magnitude in the left panels and the resulting mass-complete limit in the right panels.

    Figure \ref{fig:r-m-comp} shows the distribution of $R_{\rm e}$ and $M_{\rm *}$ in logarithmic space. The total sample is divided into quiescent and star-forming galaxies in five redshift bins ($\vartriangle z$ = 0.1). The black dotted lines show the 90\% mass-complete limit. In each redshift bin, we use blue numbers to indicate the quantity of mass-complete subsamples and red numbers to indicate the total number of samples within that bin.

    \vspace{10pt}
    \section{Size-Mass Relation} \label{sec:size-mass relation}

    As an important perspective of size-mass relation, mass growth has been discussed by many works. For example, both \cite{2014MNRAS.440..889S} and \cite{2015ApJ...814...96W} have shown that in the early phase of the galaxy evolution the dominant factor of mass growth is the continuous star formation, and mergers dominate mass growth in the later phase. The evolution of stellar mass directly affects size growth, which has also been demonstrated by many observations (e.g., \citealt{2013ApJ...771...61L,2013MNRAS.428.1715H} ).
    For instance, \cite{2010ApJ...709.1018V} demonstrate that during the redshift range of $1.4 < z < 2$, the size growth is primarily driven by star formation.
    For massive galaxies, particularly quiescent galaxies or BCGs, they exhibit a higher slope $\alpha$ in the size-mass relation, indicating that mergers have replaced star formation as the dominant factor in their mass growth ( e.g., \citealt{2009ApJ...699L.178N}). 

    However, the influence of environment on the size-mass relation has not been revealed. \cite{2012MNRAS.419.3018C} find a strong relationship between galaxy size and environment for the quiescent galaxies at $1 < z < 2$. And some studies suggest that the environment plays an important role in massive early-type galaxies at the low redshift end  (e.g., \citealt{2017ApJ...834...73Y}). But, \cite{2015MNRAS.450.1246K} compare the mass–size relations of cluster and field galaxies in the redshift range of $0.4 < z < 0.8$ and finds no significant environmental difference in it. So, how does environment play a role on size growth? Next, the impact of stellar mass and environment on the size-mass relation will be discussed in \ref{subsec:mass} and \ref{subsec:environment}, respectively.

    \subsection{The Effect of Stellar Mass} \label{subsec:mass}
 
\begin{figure*}[ht!]  
    	\includegraphics[width=\textwidth]{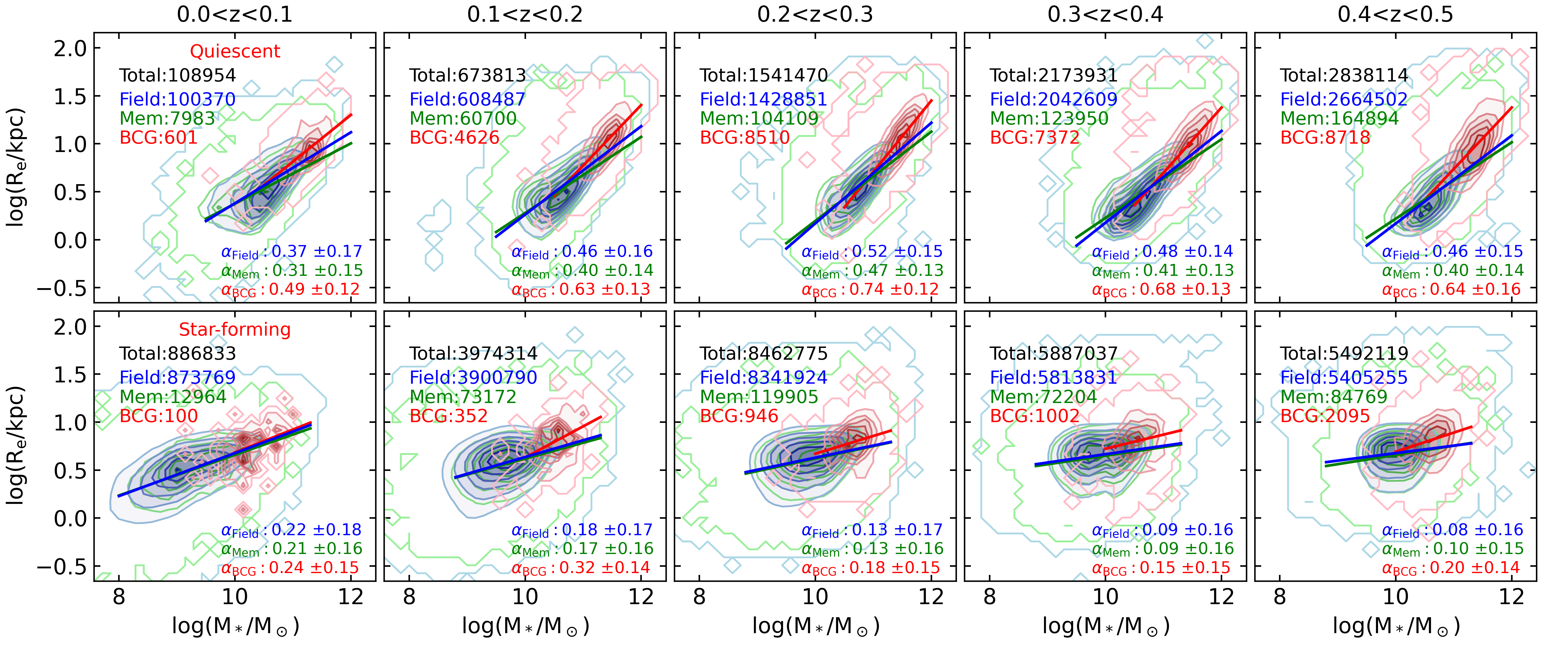}
            \caption{Contour plot of the size-mass distribution. The distribution of $R_{\rm e}$ and $M_*$ is shown in logarithmic space. The top five panels represent quiescent galaxies in different redshift bins. The contour lines in gradients of blue, green, and red colors depict the distributions of field galaxies, member galaxies, and BCGs, respectively. The legend in the top left corner shows the numbers of galaxies in different subsamples, and the slopes are shown in the lower right corner. The blue, green and red solid lines in the figure represent the least-squares fitting results of field galaxies, member galaxies, and BCGs, respectively. The bottom five panels show the distribution of star-forming galaxies.
		  \label{fig:size-mass-all}}
	\end{figure*}

    To examine the correlation between the sizes and masses of quiescent and star-forming galaxies across various redshift bins, three subsamples with distinct environmental densities are utilized. 
    We employ a robust linear model with HuberRegressor (\citealt{scikit-learn}) to fit the correlation, and Figure \ref{fig:size-mass-all} displays the results. 
    Huber loss functions possess the advantage of being less susceptible to the influence of outliers, while still taking into account their effects to a certain degree. If the absolute error of the sample is less than a certain threshold, the sample will be classified as an internal sample. We set this threshold ``epsilon" to 1.35 to achieve 95\% statistical efficiency. 
    
    The top five panels display the distributions of QGs in five different redshift bins, while the bottom five panels show the distributions of SFGs. The distributions of field galaxies, member galaxies, and BCGs are represented by gradients of blue, green, and red colors, respectively. 
    The number of galaxies in each environment is denoted by numbers of different colors, with the total number of galaxies in the bin highlighted in black, and the slopes are annotated in the lower right corner with the corresponding colors. 
    The best-fitting slopes for the QGs and SFGs in different redshift bins are presented in Table \ref{tab:4}. 

    The size-mass relations for QGs and SFGs within various environments are $R_{\rm e} \propto M_*^{0.31 \sim 0.52}$ and $R_{\rm e} \propto M_*^{0.08\sim0.22}$, respectively. Our results confirms the consistency of the size-mass relations with previous findings in the redshift range of $z<0.5$ (e.g., \citealt{2003MNRAS.343..978S,2014ApJ...788...28V,2021ApJ...921...38K}). Furthermore, we compare the slopes among the galaxies in different environments, and find that the slopes of field and member galaxies are similar. This suggests that these galaxies may have undergone similar evolutionary paths, or at least, that the denser environment does not significantly affect the size-mass relations of member galaxies. However, when comparing the slopes of field and member galaxies with those of BCGs, a noticeable steeper slope in the size-mass relation  can be seen for BCGs at $z > 0.1$, regardless of star formation status, indicating a potential environmental dependence. 
    To address this complicated phenomenon, we need to consider the mass-complete limit of the sample in order to remove the potential bias caused by mass incompleteness.

    \begin{deluxetable*}{ccccccccc}  
        \tablenum{4}
        \tablecaption{The slopes $\alpha$ of size-mass relation ($R_e \propto M_*^{\alpha}$) in different bins \label{tab:4}}
        \tabletypesize{\scriptsize}
        \tablehead{  \colhead{Sample} & \colhead{Environment} & \colhead{Type}
        & \colhead{$\rm z<0.1$} & \colhead{$\rm 0.1<z<0.2$} & \colhead{$\rm 0.2<z<0.3$} & \colhead{$\rm 0.3<z<0.4$} & \colhead{$\rm 0.4<z<0.5$} }
        \startdata
        \multirow{6}{4em}{Entire} & \multirow{2}{4em}{Field} & quiescent & 0.37 $\pm$ 0.17 & 0.46 $\pm$ 0.16 & 0.52 $\pm$ 0.15 & 0.48 $\pm$ 0.14 & 0.46 $\pm$ 0.15 \\
        &  & star-forming & 0.22 $\pm$ 0.18 & 0.18 $\pm$ 0.17 & 0.13 $\pm$ 0.17 & 0.09 $\pm$ 0.16 & 0.08 $\pm$ 0.16 \\
        \cline{2-8}
        & \multirow{2}{6em}{Member} & quiescent & 0.31 $\pm$ 0.15 & 0.40 $\pm$ 0.14 & 0.47 $\pm$ 0.13 & 0.41 $\pm$ 0.13 & 0.40 $\pm$ 0.14 \\
        &  & star-forming & 0.21 $\pm$ 0.16 & 0.17 $\pm$ 0.16 & 0.13 $\pm$ 0.16 & 0.09 $\pm$ 0.16 & 0.10 $\pm$ 0.15 \\
        \cline{2-8}
        & \multirow{2}{4em}{BCG} & quiescent & 0.49 $\pm$ 0.12 & 0.63 $\pm$ 0.13 & 0.74 $\pm$ 0.12 & 0.68 $\pm$ 0.13 & 0.64 $\pm$ 0.16 \\
        &  & star-forming & 0.24 $\pm$ 0.15 & 0.32 $\pm$ 0.14 & 0.18 $\pm$ 0.15 & 0.15 $\pm$ 0.15 & 0.20 $\pm$ 0.14 \\
        \hline
        \multirow{6}{4em}{Mass-complete} &  \multirow{2}{4em}{Field} & quiescent & 0.51 $\pm$ 0.16 & 0.60 $\pm$ 0.15 & 0.64 $\pm$ 0.13 & 0.61 $\pm$ 0.13 & 0.58 $\pm$ 0.14 \\
        &  & star-forming & 0.22 $\pm$ 0.17 & 0.17 $\pm$ 0.17 & 0.13 $\pm$ 0.16 & 0.10 $\pm$ 0.16 & 0.12 $\pm$ 0.16 \\
        \cline{2-8}
        & \multirow{2}{6em}{Member} & quiescent & 0.51 $\pm$ 0.15 & 0.57 $\pm$ 0.13 & 0.63 $\pm$ 0.12 & 0.55 $\pm$ 0.13 & 0.53 $\pm$ 0.14 \\
        &  & star-forming & 0.21 $\pm$ 0.15 & 0.16 $\pm$ 0.15 & 0.14 $\pm$ 0.15 & 0.10 $\pm$ 0.15 & 0.15 $\pm$ 0.15 \\
        \cline{2-8}
        & \multirow{2}{4em}{BCG} & quiescent & 0.50 $\pm$ 0.12 & 0.63 $\pm$ 0.13 & 0.75 $\pm$ 0.12 & 0.68 $\pm$ 0.13 & 0.64 $\pm$ 0.16 \\
        &  & star-forming & 0.24 $\pm$ 0.15 & 0.32 $\pm$ 0.14 & 0.18 $\pm$ 0.15 & 0.14 $\pm$ 0.15 & 0.21 $\pm$ 0.14 \\
        \hline
        \multirow{6}{4em}{Mass-matched} & \multirow{2}{4em}{Field} & quiescent & 0.51 $\pm$ 0.13 & 0.65 $\pm$ 0.13 & 0.69 $\pm$ 0.12 & 0.66 $\pm$ 0.12 & 0.64 $\pm$ 0.15 \\
        &  & star-forming & 0.27 $\pm$ 0.15 & 0.27 $\pm$ 0.14 & 0.21 $\pm$ 0.15 & 0.18 $\pm$ 0.16 & 0.24 $\pm$ 0.15 \\
        \cline{2-8}
        & \multirow{2}{6em}{Member} & quiescent & 0.53 $\pm$ 0.13 & 0.67 $\pm$ 0.12 & 0.72 $\pm$ 0.11 & 0.65 $\pm$ 0.12 & 0.64 $\pm$ 0.14 \\
        &  & star-forming & 0.29 $\pm$ 0.13 & 0.29 $\pm$ 0.13 & 0.23 $\pm$ 0.14 & 0.18 $\pm$ 0.15 & 0.29 $\pm$ 0.14 \\
        \cline{2-8}
        & \multirow{2}{4em}{BCG} & quiescent & 0.44 $\pm$ 0.12 & 0.58 $\pm$ 0.13 & 0.67 $\pm$ 0.12 & 0.62 $\pm$ 0.14 & 0.61 $\pm$ 0.16 \\
        &  & star-forming & 0.24 $\pm$ 0.15 & 0.32 $\pm$ 0.14 & 0.18 $\pm$ 0.15 & 0.13 $\pm$ 0.15 & 0.21 $\pm$ 0.14 \\
        \enddata	
    \end{deluxetable*}

\begin{figure*}[ht!]  
    	\includegraphics[width=\textwidth]{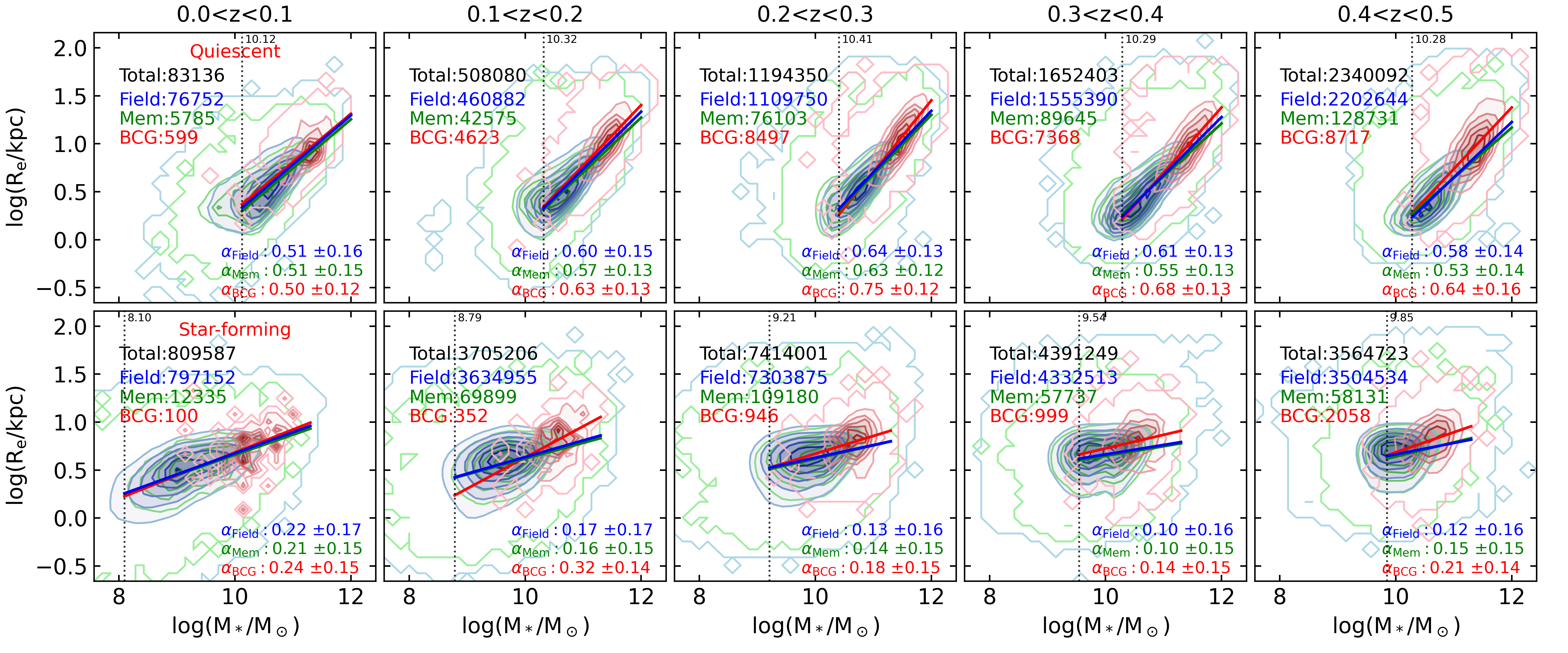}
            \caption{Contour plot of the size-mass distribution with a 90\% completeness mass limit. The distribution of $R_{\rm e}$ and $M_*$ is shown in logarithmic space. For the purpose of easy comparison, the contour lines in different colors show the distribution of the total sample. The size-mass relations and legends display the information of the mass-complete subsamples. The black dotted lines in the figure show the mass limit for the 90\% completeness sample, and the black number next to the dotted line indicates the corresponding mass-complete limit. The legend in the top left corner shows the numbers of galaxies above the mass limit in different subsamples. This is the main difference from Figure \ref{fig:size-mass-all}.
		  \label{fig:size-mass-comp}}
	\end{figure*}

    Considering the completeness of stellar mass, we fit the size-mass relations with the galaxies above the mass-complete limits listed in Table \ref{tab:3}. 
    Figure \ref{fig:size-mass-comp} presents the results of our analysis with the subsample above mass-complete limits. The best-fitting slopes of size-mass relations for all subsamples are given in Table \ref{tab:4}.
    Compared with Figure \ref{fig:size-mass-all},  field and member QGs have steeper size-mass relations, which leads to slighter difference in slope for the QGs within various environments.   
    The difference in slopes between star-forming BCGs and field/member galaxies at $z > 0.1$ persists noteworthy even with the mass-complete limit subsample.  
    
\begin{figure*}[ht!]  
    	\includegraphics[width=\textwidth]{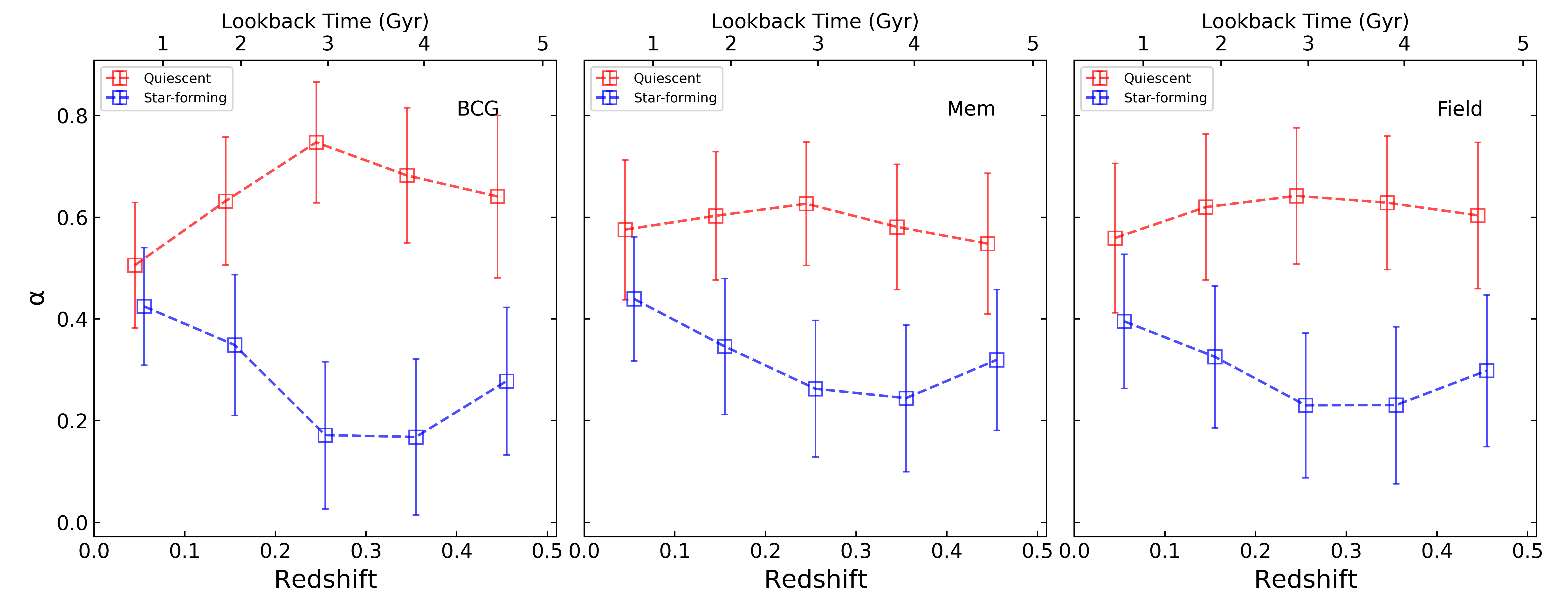}
            \caption{The slope ($\alpha$) as a function of redshift ($z$) is depicted in the figure. The slope of the size-mass relation is fitted by the sample with stellar mass greater than 10.41. We use three panels from left to right to represent the slopes of the subsamples under different environmental densities: BCG, Mem, and Field. The slopes are represented by colored squares, with red indicating quiescent galaxies and blue representing star-forming galaxies; the error bars denote the mean absolute error. Dotted lines connect the slopes across different redshift bins.
		  \label{fig:slope_mass_10.41}}
	\end{figure*}

    The slope variations with redshift are shown in Figure \ref{fig:slope_mass_10.41}. To mitigate potential effects stemming from mass incompleteness, we fit the size-mass relation for massive galaxies with log$(M_*/M_\odot) > 10.41$. Different colored squares represent the slopes of different galaxy types, while error bars denote the mean absolute errors. For quiescent galaxies, we find the highest slope in the redshift bin of $0.2 <z<0.3$, particularly pronounced for the BCGs. In the case of SFGs, steeper slopes are found at lower redshifts, whereas the slopes are lower at higher redshifts. Our study reveals different trends of redshift evolution of size-mass relation between QGs and SFGs.
    In general, the slope difference between the QGs and SFGs at $0.1<z<0.5$ is significant for various environments, and it tends to be larger at higher redshifts. 
    
    However, it remains uncertain what exactly causes the discrepancy in slope between BCGs and field/member galaxies, and it is hard to definitively attribute the difference in size-mass relation to the variations in either mass distribution (namely, subsamples in different environments are composed of galaxies of varying masses) or environmental effect. For eliminating the impact of mass distribution on size-mass relation, we will apply a mass-matching method for unveiling the net effect of environment in Section \ref{subsec:environment}.

    \subsection{The effect of environment \label{subsec:environment}}

    Whether the environment affects the size-mass relation is still under debate. At the low redshift end ($z < 0.2$), some studies show that the environment has no influence on the size-mass relation (e.g., \citealt{2010MNRAS.402..282M,2013ApJ...779...29H}). However, some people hold the opposite view. For example, \cite{2017ApJ...834...73Y} studied the massive early-type galaxies with a redshift range of $0.1 \leqslant z < 0.15$, and believed that the environment has a significant impact on the size-mass relation when the mass is greater than $10^{11.2} M_\odot$. In addition, at high redshift end, most studies show that the environment is relevant to the size-mass relation (e.g., \citealt{2013MNRAS.435..207L,2014MNRAS.441..203D}). But all of these studies are based on small samples and different definitions of environment. DESI Legacy Imaging surveys make possible for us to construct large subsamples of QGs and SFGs within different environments (namely, field galaxies, member galaxies of clusters, and BCGs).

    \begin{deluxetable*}{cccccccccc}  
        \tablenum{5}
        \tablecaption{Subsamples size and the probabilities of K-S tests for the $\rm M_*$-distributions and z-distributions \label{tab:5}}
        \tablewidth{0pt}
        \tabletypesize{\normalsize}
        \tablehead{ \colhead{Redshift} & \multicolumn{3}{c}{Galaxy number} & {} & \multicolumn{2}{c}{$\rm M_*$-distribution} & {} & \multicolumn{2}{c}{z-distribution} \\
        \cline{2-4}
        \cline{6-7}
        \cline{9-10}
        \d{Range} & \colhead{BCGs} & \colhead{Member} & \colhead{Field} & {} & \colhead{$P_{\rm M-B}$} & \colhead{$P_{\rm F-B}$} & {} & \colhead{$P_{\rm M-B}$} & \colhead{$P_{\rm F-B}$} }
        \startdata
    	$\rm 0-0.1$ & 656 & 545 & 655 &  & 0.999 & 0.981 &  & 0.810 & 0.775
    	\\
    	$\rm 0.1-0.2$ & 3,874 & 3,564 & 3,878 &  & 0.999 & 0.982 &  & 0.832 & 0.937
            \\
    	$\rm 0.2-0.3$ & 6,207 & 6,681 & 6,207 &  & 0.786 & 0.994 &  & 0.812 & 0.944
    	\\
    	$\rm 0.3-0.4$ & 6,561 & 6,470 & 6,555 &  & 0.998 & 0.996 &  & 0.998 & 0.977
    	\\
    	$\rm 0.4-0.5$ & 9,014 & 9,052 & 9,017 &  & 0.915 & 0.992 &  & 0.999 & 0.966
    	\\
        \enddata
        \tablecomments{``$P_{\rm M-B}$" represents the P-values of K-S test between member galaxies and BCGs; ``$P_{\rm F-B}$" represents the P-values between field galaxies and BCGs.}
    \end{deluxetable*}


    In order to eliminate the possible impact of mass and redshift distributions on size-mass relation, we build the mass- and redshift-matched subsamples of BCGs, member galaxies, and field galaxies at $z<0.5$. Since the number of BCGs is minimum, we try to match the field and member galaxies with similar redshift and mass distributions on the basis of BCG subsample.
    Following the mass-matching method described in \cite{2019ApJ...884..172G}, for each BCG, one matching galaxy is selected from both the field galaxies and member galaxies, respectively. The selected galaxies exhibit the minimum value of $ d = \sqrt{(z_{\rm i} - z_0)^2 + (\log M_{\rm *,i} - \log M_{\rm *,0})^2 } $, where ($z_0, M_{\rm *,0}$) represents the position of the given BCG in the redshift-mass space, and ($z_{\rm i}, M_{\rm *,i}$) represents the location of the galaxies to be matched. For ensuring a robust redshift- and mass-matching, the upper limit of $d$ is set to be 0.04. Each BCG is matched with a similar galaxy in both mass and redshift space. The BCGs that fail to find counterpart galaxies in the above matching process are discarded.
    As a result, we obtain the mass-matched subsamples of BCGs, field galaxies and member galaxies with similar redshift- and mass-distributions. 
    Finally,  three mass-matched subsamples including 26,312 BCGs, 26,312 massive member galaxies, and 26,312 field galaxies are constructed. The Kolmogorov–Smirnov (K-S) tests are performed to check the similarities in mass- and redshift-distributions. Table \ref{tab:5} lists the sample size and the K-S probabilities ($P$). All of the $P$-values are much larger than the threshold of 0.05, which ensures the similarities in $M_*$- and $z$-distributions for the mass-matched samples. 
    
\begin{figure*}[ht!]  
            \includegraphics[width=\textwidth]{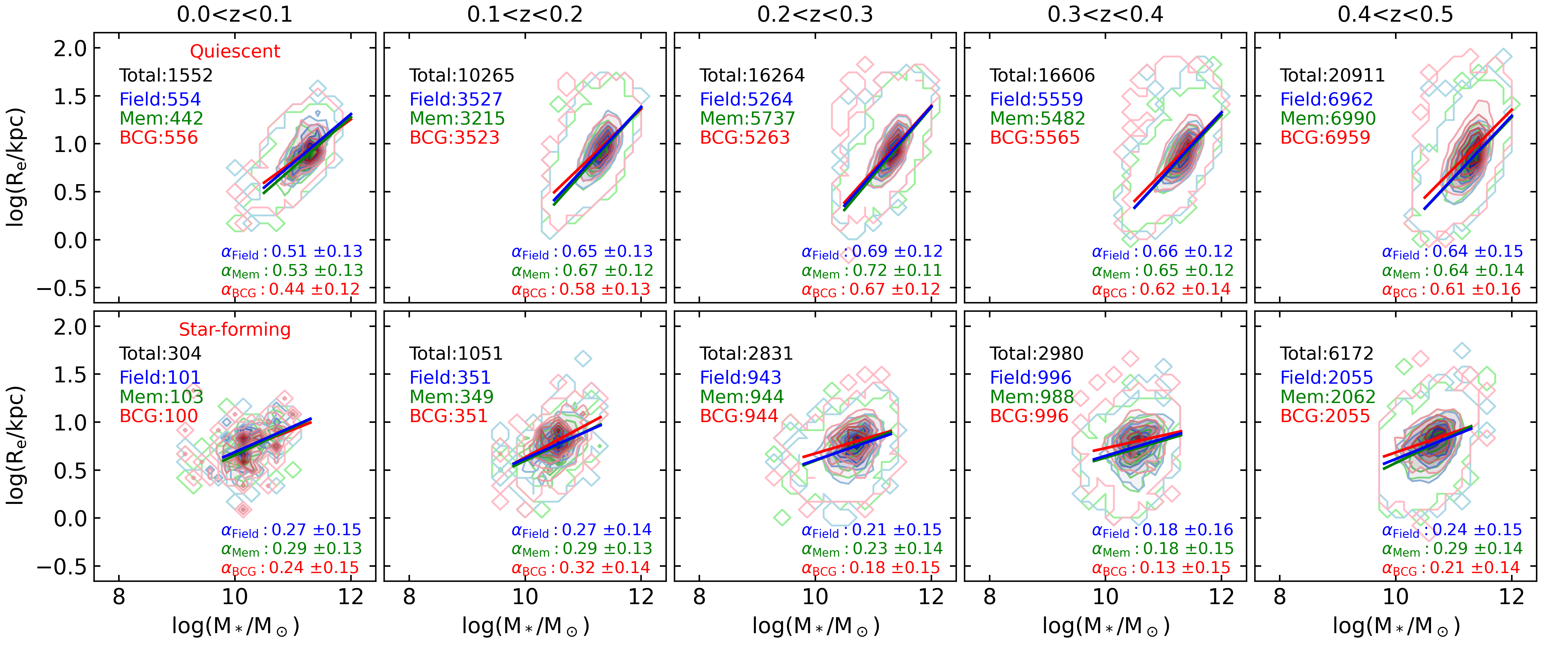}
            \caption{Contour plot of the size-mass distribution with the mass-matched subsamples. The distribution of $R_{\rm e}$ and $M_*$ is shown in logarithmic space. 
            \label{fig:size-mass-match}}
	\end{figure*}

    Figure \ref{fig:size-mass-match} shows the size-mass relations for the QGs and SFGs in mass-matched subsamples of BCGs, member galaxies and field galaxies. The best-fitting slopes of size-mass relations for mass-matched subsamples are listed in Table \ref{tab:4}.
    Compared with Figure \ref{fig:size-mass-comp}, both field and member galaxies have steeper size-mass relations, while quiescent BCGs tend to have flatter size-mass relations. The QGs in fields and clusters show obvious steepening trends in size-mass relations for various redshift bins than the SFGs. In Figure \ref{fig:size-mass-match}, we observed a clear difference in the intercept of the size-mass relation between the quiescent BCGs and field/member galaxies at $0.4 < z < 0.5$. 
    However, when the uncertainties of slopes are taken into account, the slope differences among the mass-matched subsamples of field galaxies, member galaxies of clusters, and BCGs are negligible for both QGs and SFGs.
    This suggests that the significant discrepancy in slope between BCGs and field/member galaxies unveiled in Figure \ref{fig:size-mass-comp} is mainly because BCGs are dominated by massive galaxies covering the high end of mass distribution. After eliminating the effect of mass distribution, environment seems to have little effect on size-mass relations of QGs and SFGs. In other words, mass distribution has been affected by environment, but the ultimate galaxy size is mainly determined by its stellar mass.
    
\begin{figure*}[ht!]  
    	\includegraphics[width=\textwidth]{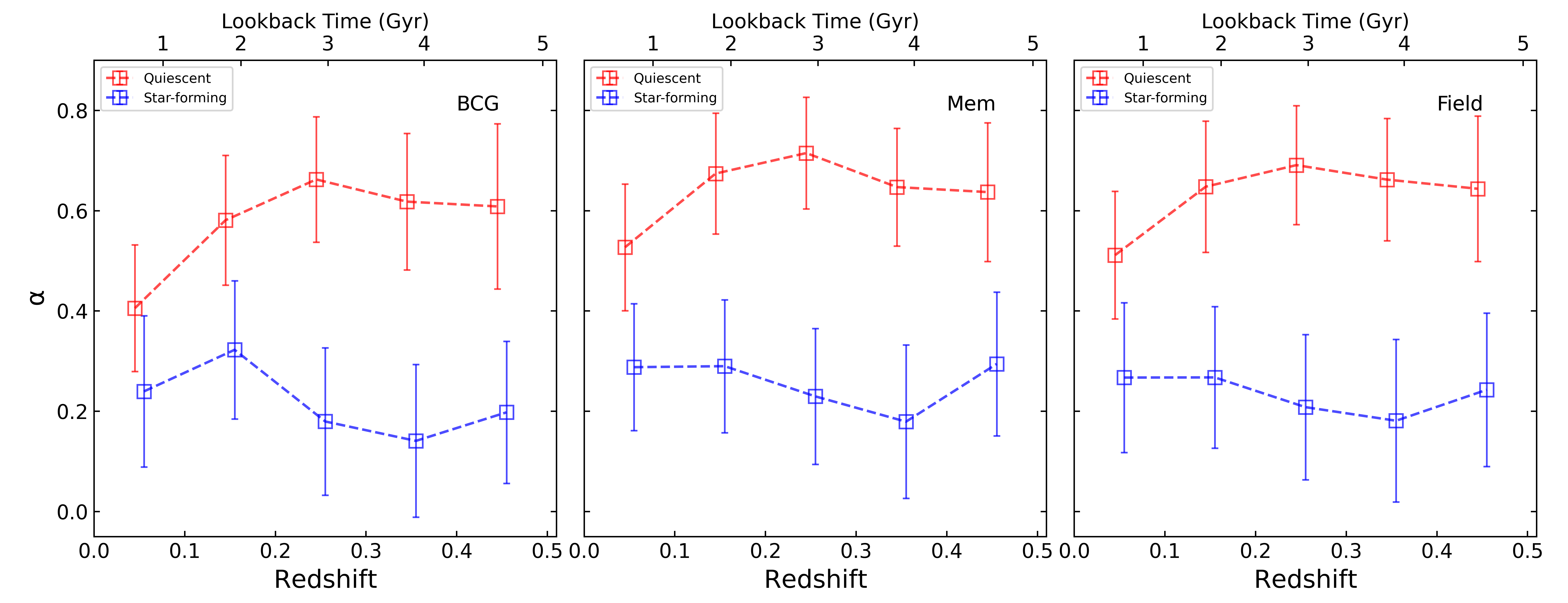}
            \caption{The slope ($\alpha$) as a function of redshift ($z$) is depicted in the figure. The slope of the size-mass relation is fitted by the mass-matched subsamples. We use three panels from left to right to represent the slopes of the subsamples under different environmental densities: BCG, Mem, and Field. The slopes are represented by colored squares, with red indicating quiescent galaxies and blue representing star-forming galaxies; the error bars denote the mean absolute error. Dotted lines connect the slopes across different redshift bins.
		  \label{fig:slope_match}}
	\end{figure*}

    The slope variations with redshift for the mass-matched subsamples are shown in Figure \ref{fig:slope_match}. Compared with Figure \ref{fig:slope_mass_10.41}, it is confirmed that the mass-matched subsamples of QGs have the highest slope at $0.2 < z < 0.3$, but the trend of SFGs having a higher slope at $z<0.2$ seems to be alleviated, showing a more consistent slope across different redshift bins. It should be noted that the trend of slope change with redshift for the QGs and SFGs within three environments is more consistent for the mass-matched subsamples, which further confirms that the size-mass relation is independent upon the environment.

    \cite{2017ApJ...834...73Y} found that the early-type galaxies with $ M_* > 10^{11.2} M_\odot$ in denser environment tend to have larger effective radii on the basis of random sampling method, which was attributed to environmental factors. 
    Since the majority of early-type galaxies in high-density regions are quiescent, it is interesting for us to check this tendency via our subsamples of quiescent galaxies. 
    Based on our large mass-matched subsamples of quiescent BCGs, member galaxies, and field galaxies, we achieve different results. 
    It is found that the massive QGs with $M_* > 10^{11.2} M_\odot$ indeed exhibit a higher slope in their mass-size relation with respect to the entire mass-matched QG subsample. However, there is no significant slope difference among the massive BCGs, members, and field galaxies with $M_* > 10^{11.2} M_\odot$. Our finding supports that the steeper slopes in size-mass relation observed in massive galaxies across different environments can be attributed to the difference in the distribution of stellar mass. The environment does not directly influence the size-mass relation, but rather changing the distribution of stellar mass. Stellar mass seems to be the predominant factor for driving the size growth of galaxies.

    To sum up, in the star-forming epoch of galaxy evolution (i.e.,SFGs), continuous star formation leads to stellar mass growth. With the increase in stellar mass and gas consumption, star formation activity will halt when its gas supply and reservoir are destroyed. For the quiescent epoch of galaxy evolution, minor mergers become the dominant mode of mass growth in QGs  (\citealt{2013ApJ...773...43B,2015ApJ...813...23V,2007ApJ...658..710N}). This leads to a higher prevalence of massive galaxies for QGs, which is why QGs exhibit a steeper size-mass relation than SFGs (see Figure \ref{fig:size-mass-comp}). 
    In general, massive galaxies are more likely to be attracted towards the centers of high-density dark matter halos due to gravitational interactions, which provides more opportunities for galaxy mergers. 
    This makes the high density environment be occupied by more massive galaxies, which is the reason why the most of BCGs are massive QGs. 
    Our work suggests that, for QGs and SFGs, stellar mass is more fundamental factor for size growth, and  environments only affect the distribution of stellar mass.

    \section{ Discussion
    \label{sec:discussion}}
    
    \subsection{Uncertainties of stellar mass and SFR} \label{subsec:4.1}
    
    To enhance the robustness of QGs and SFGs classification, we simultaneously employ two criteria: color ($g-z$) and {\rm sSFR}. The stellar mass and SFR used in this work are derived with the LePhare software by \cite{2019ApJS..242....8Z}.
    As described in  \cite{2019ApJ...873...44C}, inferring the SFH from the galaxy SED is quite challenging due to the strong evolution of the mass-to-light ratios of stellar populations with age. 
    Generically, the SED-fitting method can recover the stellar mass at an uncertainty of about 0.2 – 0.3 dex, and even better than 0.2 dex if near-infrared bands are involved \citep{2009ApJ...701.1839M,2019ApJS..242....8Z}. 
    In our study, the robust photometric redshifts and stellar masses for galaxies are derived from both optical and near-infrared photometry.

    There is a high level of consistency in stellar mass found in \cite{2019ApJS..242....8Z} when comparing their results with MAGPHYS fitting. In addition, \cite{2023arXiv230703725K} also indicate that, despite the various SFH patterns observed in individual galaxies, the median {\rm sSFR} is consistent across all mass bins. Nevertheless, it is important to note that {\rm sSFR} is an exceptionally sensitive indicator in sampling.
    According to the findings of \cite{2017ApJ...837..170L}, the adoption of nonparametric SFHs in Prospector-$\alpha$, as opposed to the use of parametric SFHs in MAGPHYS, results in an average increase of 0.25 dex in stellar mass and a 0.1 dex decrease in SFR. Therefore, in order to valuate the possible effect due to sSFR inconsistency produced by various SFH treatments, some blue galaxies within $-11 < \log {\rm sSFR} < -10.65$ might have been misclassifed as SFGs. Number of these blue galaxies is 400,816, occupying only 1.62\% of our star-forming subsample. 
    Consequently, we have strong grounds to dismiss the influence of different fitting models on the outcomes of our research. 
    
    \subsection{Size-mass relation fitting and environment effect} \label{subsec:4.2}

    In general, changes in the mass of a galaxy will be accompanied by alterations in its size. Numerous factors impact the transformations in the mass and size of a galaxy, including accretion, activities of star formation, mergers, and environment. We employ the size-mass relation ($R_{\rm e} \propto M_*^{\alpha}$), which is a fundamental relation, to explore the size growth of galaxies in relation to stellar mass and environment, further examine the effects of environment on size growth. \cite{2003MNRAS.343..978S} apply a power-law to fit the size-mass  relations for early- and late-type galaxies, respectively. 
    A difference in the size-mass relations is found between the low- and high-mass ends of late-type galaxies. It shows a steeper relation $R \propto M_*^{0.4}$ at $M_* >  M_0 = 10^{10.6} M_\odot$, and a shallower relation $R \propto M_*^{0.15}$ below $M_0$. They also find that the relation is significantly steeper for early-type galaxies ($R \propto M_*^{0.55}$). \cite{2014ApJ...788...28V} confirm that the size-mass relation obey a single power law for the early- and late-type galaxies over a large redshift range of $0 < z < 3$. A growing body of research suggests that the relationship between size and mass varies across different mass ranges, deviating from a single power law. \citet{2019ApJ...872L..13M} fit the size-mass relation of all types of galaxies (i.e., QGs and SFGs) with a broken power-law up to $z \thicksim 3$. 
    Analogously, \cite{2021ApJ...921...38K} use a broken power-law to fit the size-mass relation for QGs and SFGs at $z < 1$, respectively. They demonstrate that the size-mass relations of both populations with a clear change of slopes at a pivot stellar mass $M_{\rm p}$ ($M_{\rm p} \thicksim 10^{10.2-10.6} M_\odot$ for QGs, $M_{\rm p} \thicksim 10^{10.7} M_\odot$ for SFGs). It may imply that stellar mass is a dominant factor in size growth.

    In this paper, we mainly focus on comparing the difference in size-mass relation for the QGs and SFGs at $z<0.5$  within various environments, and try to observe the fundamental factor driving the change of the slope. The majority of BCGs commonly have higher stellar masses, resulting in the BCG subsample covers a narrower mass range.  It is hard to perform a broken power-law to fit the size-mass relations for mass-matched subsamples. Therefore, we adopt single power-law in our fittings.

    Using single power-law fitting, the recent work by \cite{2023A&A...670A..95A} studied the size-mass relation for the galaxies in 15 clusters at $1.4<z<2.8$ from the CARLA survey. It is found that the size-mass relation does not evolve much from $z\sim 2$ to the present, and their sizes evolve in a similar way for cluster and field galaxies. Moreover, both BCGs and satellite galaxies are found to lie on the same size-mass relation. Similarly, \cite{2023A&A...669A.131S} compare the sizes of field galaxies at $1.4<z<1.7$ with the data from massive clusters in the South Pole Telescope Sunyaev Zel'dovich effect survey, and no significant differences in size are revealed between these two different environments. 
    It is interesting that the massive galaxies with $M_* \sim 10^{11} M_{\odot}$  observed by the James Webb Space Telescope (JWST) in near infrared band appear to be more compact than their rest-frame optical light profiles at cosmic noon on average (\citealt{2022ApJ...937L..33S}).
    It is worth noting that their studies primarily focus on the galaxies at high redshifts, whereas our current work takes use of the galaxies at $z < 0.5$. 
    The environmental effect on the star formation quenching tends to be more significant at $z<0.5$ (\citealt{2010ApJ...721..193P}). The large sample of the $z<0.5$ galaxies from the DESI Legacy Imaging Surveys has great advantages in studying the environment effects on size growth.   Our results suggest that stellar mass distribution dominates the size-mass relation, and environmental factors (such as tidal stripping, ram pressure stripping, and harassment) only change their star formation histories. The observed sizes tend to be more closely relative to the accumulated stellar masses, whether they are in field or clusters, and whether they are central or satellite galaxies.

    \subsection{Redshift evolution of size-mass relation} \label{subsec:4.3}

    \begin{figure*}[ht!]  
        \includegraphics[width=\textwidth]{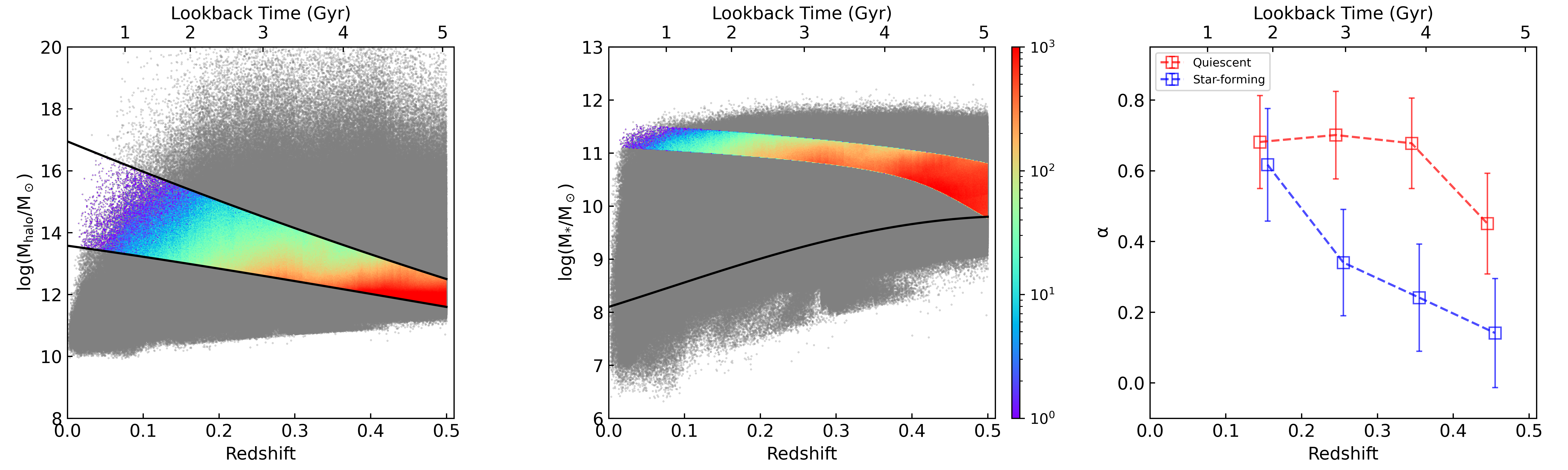}
            \caption{Left: the distribution of halo mass and redshift for all the galaxies selected in our work. The colored points denote the galaxies in our growth-tracking subsample; the solid black lines in this panel represent the MAHs for the upper (12.5) and lower (11.6) halo mass limits at $z=0.5$. Middle: the distribution of stellar mass and redshift for all the galaxies selected in our work. The color bar represents the number of galaxies; the black line represents the mass-complete limit, which is fitted by interpolation. Right: the slope distribution of the size-mass relation for the subsample of growth-tracking galaxies at $0.1 < z < 0.5$. The slopes are represented by colored squares, with red indicating QGs and blue representing SFGs; the error bars denote the mean absolute error.
          \label{fig:figure_10}}
    \end{figure*}

    In Section \ref{subsec:mass} and \ref{subsec:environment}, we demonstrate the evolution of the slope for QGs and SFGs in different environments with both mass-complete and mass-matched subsamples, respectively. 
    When considering uncertainties and mass distribution, a similar slope is found for the SFGs in various environments, and a steeper slope is found for the QGs, particularly at $z>0.1$.
    Generally, as galaxies transition from SFGs to QGs, the mass growth pattern also undergoes changes, which, in turn, results in alternations in the slope. However, in the context of galaxy evolution, the galaxies with lower masses at higher redshifts should be the progenitors of those at present ($z=0$). It is interesting to discuss the redshift evolution of the size-mass relation on the basis of the galaxy samples at various redshifts with some possible evolutionary link.

    In order to minimize the influence of progenitor bias,  we adopt the model of mass accretion histories (MAH), given by \cite{2009MNRAS.398.1858M} on the basis of dark matter halos from the Millennium simulation, which can be formulated with $ M(z) = M_0 (1 + z)^\beta e^{-\gamma z}$, where $M_0$ represents the halo mass at $z = 0$, and $\beta$ and $\gamma$ are fitting parameters.
    By using the MAH model mentioned above, \cite{2017ApJ...843...28M} took the cluster halo mass at high redshift to track the expected growth trajectory of clusters at lower redshifts. 
    Similarly, we take the halo mass region, $\log M_{\rm halo}/M_{\odot} \thicksim 11.6-12.5$, as the galaxy-size halos at $z = 0.5$ to trace the corresponding ranges of galaxy halo mass at lower redshifts. 
    Based on the known stellar mass, the corresponding halo mass of a galaxy can be estimated by the stellar-to-halo mass relation (SHMR) that is recently established by 
    \cite{2022A&A...664A..61S} with the COSMOS2020 catalog. For the galaxies at $z<0.5$, the SHMR can be modeled as
    $$
        \log M_{\rm halo} = \log M_1 + \beta \log \frac{M_*}{M_{*,0}} + \frac{ {({M_*}/{M_{*,0}})}^{\delta}}{1 +  {({M_*}/{M_{*,0}})}^{-\gamma}} - 0.5,
    $$
    where all of $\log M_1$, $\log M_{*,0}$, $\beta$, $\delta$ and $\gamma$ are the best-fitting parameters given in \cite{2022A&A...664A..61S}.
    
    The upper and lower limits for halo mass ($\log M_{\rm halo}/M_{\odot}$) at $z=0.5$ are set to be 11.6 and 12.5 (for a typical galaxy-sized halo), respectively. The stellar mass ($\log M_*/M_{\odot}$) corresponding to the lower limit of halo mass is 9.8, which is the lower limit of stellar mass for the mass-complete sample at 0.4 $<z<$ 0.5. The derived upper limit of stellar mass at $z=0.5$ is $\log M_*/M_{\odot}=10.8$. 
    
    As depicted in Figure \ref{fig:figure_10}, the left panel illustrates the distribution of halo mass across different redshifts. Colored points represent our growth-tracking subsample, while the solid black lines in this panel represent the MAHs for the upper and lower mass limits.
    In the middle panel, one can observe the distribution of stellar mass for our subsample, with the color bar indicating the number of galaxies.
    The right panel displays the distribution of slopes for the growth-tracking subsample. The BCGs have been excluded from this subsample due to their complex evolutionary processes. Due to the limited sample size and larger slope uncertainty at $z < 0.1$, we primarily focus on the slope relation for $0.1 < z < 0.5$.
    As shown in the right panel, the SFGs exhibit an increase in their slope at lower redshifts, suggesting a change in their size growth pattern. In contrast, QGs demonstrate a relatively stable slope within the range of $0.1 < z < 0.4$, indicating a stable size growth pattern.
    
    We would like to emphasize that the evolutionary link between the galaxies at different redshifts is strongly dependent upon the halo growth model based on the simulations only. Moreover, some QGs at $z=0$ might have been evolved from the SFGs at higher redshifts, and the progenitors of some SFGs might be quiescent.  For simplicity we neglect the cases that the type of progenitors changes from $z=0.5$ to $0.1$. Therefore, this strategy to create galaxy samples with evolutionarily links should be further validated through a detailed comparison between simulations and observations.

    \section{Conclusion} \label{sec:conclusion}

    In this paper we explore the size-mass relations of quiescent and star-forming galaxies in three distinct environments (i.e., BCGs, cluster galaxies, field galaxies), using a large sample of 32,039,360 galaxies with $r < 22$ and $z < 0.5$ selected from the DESI Legacy Imaging Surveys. Based on the color ($g$ - $z$) and {\rm sSFR} criteria, these galaxies are divided into QGs and SFGs. The BCGs are found to be dominated by QGs. The difference in size-mass relation for the galaxies within three environments has been investigated, and the main conclusions are the followings:
    
    \begin{enumerate}[noitemsep, topsep=0pt]
        \item Regardless of galaxy environment, QGs exhibit steeper size-mass relation compared with SFGs, based on both mass-complete and mass-matched subsamples. The slope difference between QGs and SFGs is significant at $0.1 < z < 0.5$ for various environments.
        
        \item The galaxies with greater stellar masses demonstrate steeper size-mass relation. For mass-complete subsamples, BCGs show the highest slope in the size-mass relation when compared to member and field galaxies. No significant difference in slopes is observed between cluster members and field galaxies. 
        
        \item The slope variations of the size-mass relation with redshift exhibit distinct characteristics for QGs and SFGs. 
        The slope of QGs reaches its maximum at $0.2 < z < 0.3$, while the slope of SFGs is higher at $z < 0.2$.
        Compared to the mass-complete subsamples,  the mass-matched subsamples exhibit a more consistent trend of slope variation with redshift for the QGs and SFGs within three environments.
        
        \item Based on the mass-matched subsamples of QGs and SFGs, no significant slope difference for the size-mass relation is found among BCGs, cluster members, and field galaxies. It suggests that, though mass growth mode and mass distribution for QGs and SFGs can be affected by galaxy environment, stellar mass is the most fundamental factor driving the size evolution at $z < 0.5$. 
        
    \end{enumerate}
    \vspace{10pt}

    This research is supported by National Key R\&D Program of China No. 2023YFA1607800 and the National Natural Science Foundation of China (NSFC, Nos. 12273013, 12373010, 12120101003 and 11890691). Y.G. acknowledges the support from China Postdoctoral Science Foundation (2020M681281) and Shanghai Postdoctoral Excellence Program (2020218). H. Z. acknowledges the supports from the National Key R\&D Program of China (Grant No. 2022YFA1602902) and China Manned Space Project (Nos. CMS-CSST-2021-A02 and CMS-CSST-2021-A04).

    The DESI Legacy Imaging Surveys consist of three individual and complementary projects: the Dark Energy Camera Legacy Survey (DECaLS), the Beijing-Arizona Sky Survey (BASS), and the Mayall z-band Legacy Survey (MzLS). DECaLS, BASS and MzLS together include data obtained, respectively, at the Blanco telescope, Cerro Tololo Inter-American Observatory, NSF's NOIRLab; the Bok telescope, Steward Observatory, University of Arizona; and the Mayall telescope, Kitt Peak National Observatory, NOIRLab. NOIRLab is operated by the Association of Universities for Research in Astronomy (AURA) under a cooperative agreement with the National Science Foundation. Pipeline processing and analyses of the data were supported by NOIRLab and the Lawrence Berkeley National Laboratory (LBNL). Legacy Surveys also uses data products from the Near-Earth Object Wide-field Infrared Survey Explorer (NEOWISE), a project of the Jet Propulsion Laboratory/California Institute of Technology, funded by the National Aeronautics and Space Administration. Legacy Surveys was supported by: the Director, Office of Science, Office of High Energy Physics of the U.S. Department of Energy; the National Energy Research Scientific Computing Center, a DOE Office of Science User Facility; the U.S. National Science Foundation, Division of Astronomical Sciences; the National Astronomical Observatories of China, the Chinese Academy of Sciences and the Chinese National Natural Science Foundation. LBNL is managed by the Regents of the University of California under contract to the U.S. Department of Energy.


    \clearpage
    \appendix 
    \section{The accuracy of photometric redshift} \label{sec:appendix}

    We adopt three parameters to evaluate the quality of photometric redshift at $0 < z < 0.5 $ as follows: 
	
	\begin{enumerate}
            \item Bias ($\vartriangle_{\rm NMAD}$): the median of systematic deviation between the spectroscopic ($z_{\rm spec}$) and photometric redshifts ($z_{\rm photo}$),
		\begin{small}
			\begin{eqnarray*}
                \vartriangle_{\rm NMAD} = \rm median\left( \frac{\mathit{z_{\rm photo}}- \mathit{z_{\rm spec}}}{1 + \mathit{z_{\rm spec}}}  \right).
			\end{eqnarray*}
		\end{small}				 	
            \item Dispersion ( $\sigma_{\vartriangle_{\rm NMAD}} $): the dispersion of $ \vartriangle_{\rm NMAD}$, and the median $\vartriangle\!z (= z_{\rm photo} - z_{\rm spec}$) is used to calculate normal distribution of median absolute deviation,
		\begin{small}
			\begin{eqnarray*}
                \! \sigma_{\vartriangle_{\rm NMAD}}\! = 1.48 \times \rm \nonumber median\left( \frac{|\vartriangle\!\mathit{z} - median(\vartriangle\!\mathit{z})| }{1 + \mathit{z_{\rm spec}}} \right). 
			\end{eqnarray*}
		\end{small}
            \item Outlier rate ($\eta_{0.15}$): the fraction of galaxies exhibiting significant deviations between their photometric redshifts and spectroscopic redshifts, satisfying 
		\begin{small}
			\begin{eqnarray*}
                \rm \eta = \frac{|\mathit{z_{\rm photo}} - \mathit{z_{\rm spec}}|}{1+ \mathit{z_{\rm spec}}} > 0.15.
			\end{eqnarray*}
		\end{small}	
	\end{enumerate}

        \begin{figure}[ht!]  
            \centering
            \includegraphics[width=0.35\columnwidth]{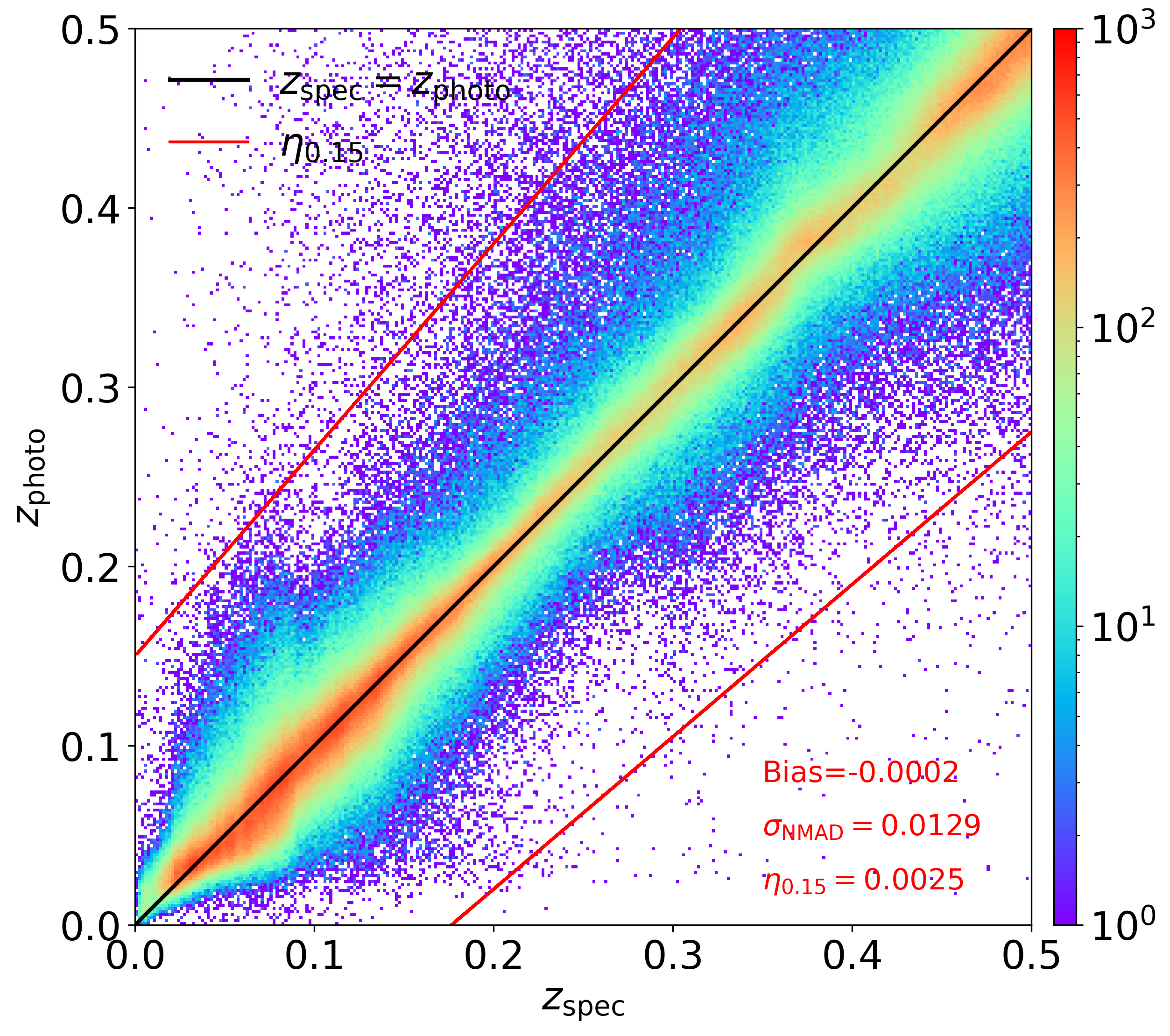} 
            \caption{Photometric redshift ($ z_{\rm photo}$) as a function of spectroscopic redshift ($z_{\rm spec}$). The color gradients in the figure represent the density of the galaxies. The black solid line in the figure indicates that $z_{\rm spec}$ = $z_{\rm photo}$, the red solid lines show the outlier rate of photometric redshifts with $\eta > 0.15$. The value of the three parameters (bias, dispersion and outlier rate) are marked in red at the lower right corner. \label{fig:z-z}}
        \end{figure}
        
\begin{figure}[ht!]  
            \centering
            \includegraphics[width=0.85\columnwidth]{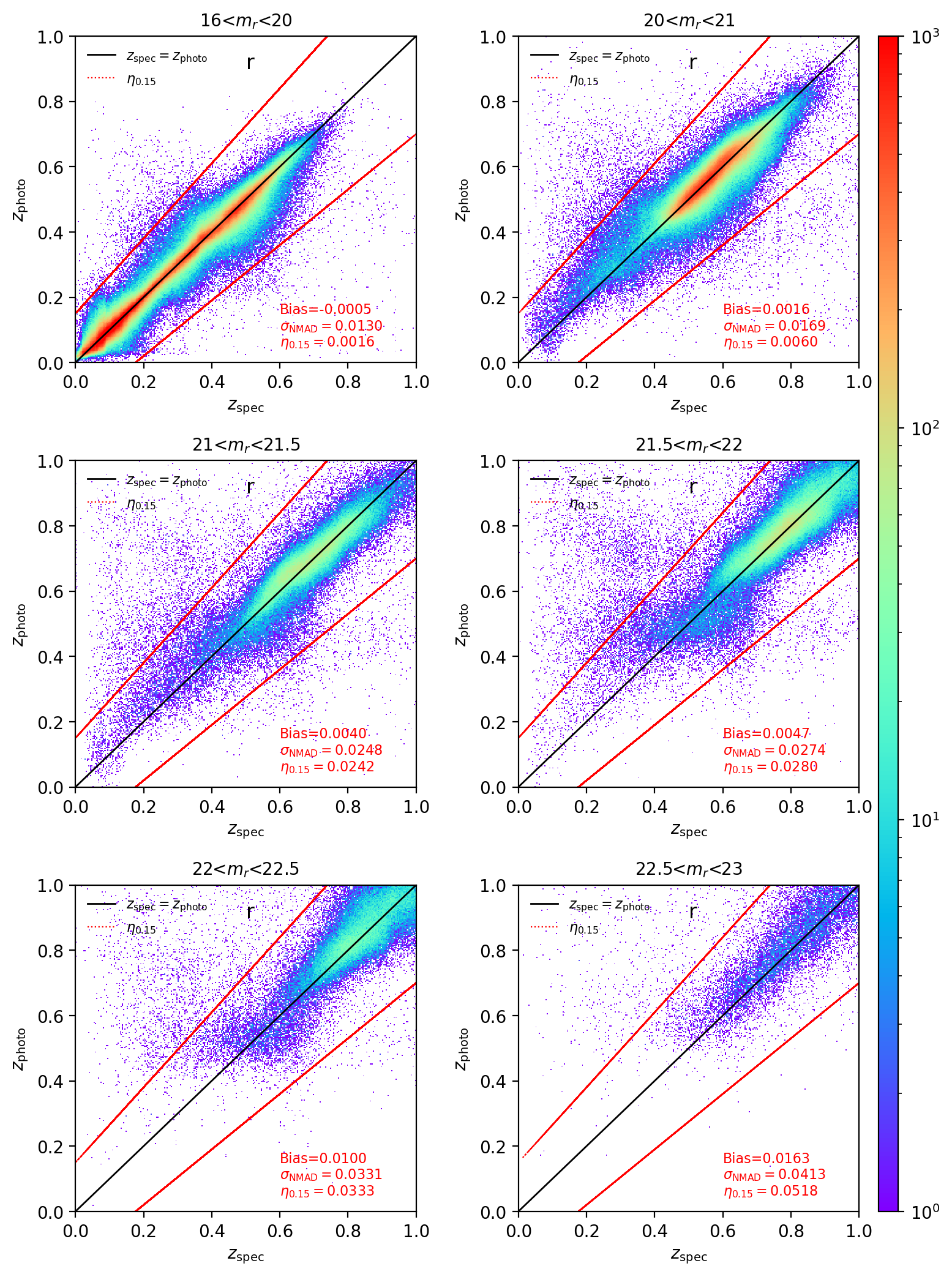} 
            \caption{Photometric redshifts v.s. spectroscopic redshifts in six  different $r$-band magnitude bins. The color gradients in the figure represent the density of the galaxies. The black solid line in the figure indicates that $z_{\rm spec}$ = $z_{\rm photo}$, the red solid lines show the outlier rate of photometric redshifts with $\eta > 0.15$. The value of the three parameters (bias, dispersion and outlier Rate) are marked in red at the lower right corner.
            \label{fig:z-z-mag}}
        \end{figure}

        The accuracy of photometric redshifts in the range of $z < 0.5$, is $\sigma_{\vartriangle_{\rm NMAD}} \approx $ 0.013, as shown in Figure \ref{fig:z-z}. Compared with the accuracy about 0.017 for the total sample in \cite{2019ApJS..242....8Z}, the photometric redshift accuracy is better at $z<0.5$. To verify the accuracy of photometric redshift as a function of magnitude, we divided the magnitude into six bins, as shown in Figure \ref{fig:z-z-mag}. 
        Three parameters (bias, dispersion and outlier rate) are labelled at the lower right corner of each panel. For the galaxies with $r > 22$, both dispersion and outlier rate exceed 0.03. 
        One possible reason is the lack of spectroscopic data as training sample at the faint end (see also Figure 2 of  \citealt{2022RAA....22f5001Z}). 
        The large uncertainty of redshift would enhance the contamination in membership of clusters. To ensure the purity of cluster members, we set the $r$-band magnitude limit to 22 in further sample selection.

    \newpage
    \begin{singlespace}
        \bibliography{references}
        \addcontentsline{toc}{chapter}{REFERENCES}
    \end{singlespace}
    
\end{document}